\documentclass[12pt]{article}

\usepackage{graphicx}
\begin{document}

%New Commands
\newcommand{\siml}{\stackrel{<}{\sim}}
\newcommand{\simg}{\stackrel{>}{\sim}}
\newcommand{\lleq}{\stackrel{<}{=}}

\baselineskip=1.333\baselineskip
%PRE
%\baselineskip=2.0\baselineskip

%\draft
%\baselineskip=0.5\baselineskip

%
\begin{center}
{\large\bf
Stationary and dynamical properties of information entropies \\
in nonextensive systems
%the coupled Langevin model subjected to multiplicative noise
} 
%\footnote{E-print: arXiv:0711.3923}
\end{center}

\begin{center}
Hideo Hasegawa
\footnote{E-mail address:  hideohasegawa@goo.jp}
\end{center}

\begin{center}
{\it Department of Physics, Tokyo Gakugei University  \\
Koganei, Tokyo 184-8501, Japan}
\end{center}
\begin{center}
%{\rm (Jan. 11, 2005)}
({\today})
\end{center}
%\maketitle
\thispagestyle{myheadings}

\begin{abstract}
The Tsallis entropy and Fisher information entropy (matrix)
are very important quantities expressing information measures
in nonextensive systems. Stationary and dynamical
properties of the information entropies have been investigated 
in the $N$-unit coupled Langevin model
subjected to additive and multiplicative white noise,
which is one of typical nonextensive systems.
We have made detailed, analytical and numerical study 
on the dependence of the stationary-state entropies on 
additive and multiplicative noise, external inputs, couplings 
and number of constitutive elements ($N$).
By solving the Fokker-Planck equation (FPE) by both the proposed 
analytical scheme and the partial difference-equation method,
transient responses of the information entropies
to an input signal and an external force have been investigated. 
We have calculated the information entropies also with the use of
the probability distribution derived by
the maximum-entropy method (MEM), whose result is compared to
that obtained by the FPE. 
The Cram\'{e}r-Rao inequality is shown to be expressed
by the {\it extended} Fisher entropy, which is
different from the {\it generalized} Fisher entropy
obtained from the generalized Kullback-Leibler divergence
in conformity with the Tsallis entropy.
The effect of additive and multiplicative
{\it colored} noise on information entropies is discussed also.

\end{abstract}

%\noindent
\vspace{0.5cm}


\newpage
%\narrowtext
\section{INTRODUCTION}

In the last half century, considerable studies have been made on
the Boltzman-Gibbs-Shannon entropy and the Fisher information 
entropy (matrix), both of which play important roles 
in thermodynamics and statistical mechanics of classical and 
quantum systems \cite{Frieden98}-\cite{Goswami05}.
The entropy flux and entropy production have been investigated
in connection with the space volume contraction
\cite{Daems99}. 
In the information geometry \cite{Amari00}, the Fisher information matrix
provides us with the distance between the neighboring points 
in the Rieman space spanned by probability distributions. 
The Fisher information matrix gives
the lower bound of estimation errors in the Cram\'{e}r-Rao theorem.
In a usual system consisting of $N$ particles,
the entropy and energy are proportional to $N$ ({\it extensive}), 
and the probability distribution is given by the Gaussian distribution 
belonging to the exponential family. 

In recent year, however, many efforts have been made
for a study on {\it nonextensive} systems in which the physical 
quantity of $N$ particles is not proportional to $N$
\cite{Tsallis88,Tsallis98,Tsallis04}. 
The nonextensivity has been realized in various systems
such as a system with long-range interactions,
a small-scale system with large fluctuations in temperature
and a multi-fractal system \cite{Tsallis04,Nonext}. 
Tsallis has proposed the generalized entropy
(called the Tsallis entropy hereafter) defined by
\cite{Tsallis88,Tsallis98}
\begin{eqnarray}
S_q(t) &=& \frac{k}{(q-1)} \left(1-\int p(x,t)^q \:dx \right),
\label{eq:A1} \\
&=& - k\:\int p(x,t)^q \;\ln_q p(x,t)\:dx,
\label{eq:A2} 
\end{eqnarray}
where $q$ is the entropic index, $p(x,t)$ denotes the probability 
distribution of a state $x$ at time $t$, the Boltzman constant $k$ 
is hereafter unity and $\ln_q x$ expresses the  $q$-logarithmic function
defined by $\ln_q x \equiv(1-x^{1-q})/(q-1)$.
The Tsallis entropy accounts for the nonextensivity of the entropy 
in nonextensive systems. In the limit of $q \rightarrow 1$, 
$\ln_q x$ reduces to the normal $\ln x$ and then $S_q(t)$ agrees with
the Boltzman-Gibbs-Shannon entropy expressed by
\begin{eqnarray}
S_1(t) = - \int p(x,t) \ln p(x,t)\: dx.
\label{eq:A3}
\end{eqnarray}
The probability distribution derived by the maximum-entropy method (MEM)
with the use of the Tsallis entropy is given by non-Gaussian 
distribution \cite{Tsallis04}, which reduces to the Gaussian and 
Cauchy distributions for $q=1$ and $q=2$, respectively.

Many authors have discussed the Fisher information matrix 
in nonextensive systems \cite{Plastino95}-\cite{Masi06}.
In order to derive the {\it generalized} Fisher information matrix
{\sf G} whose components are given by \cite{Abe03}-\cite{Masi06}
\begin{eqnarray}
g_{ij} &=& g_{ji}
=q \int p(x) \left( \frac{\partial \ln p(x)}{\partial \theta_i}\right)
\left( \frac{\partial \ln p(x)}{\partial \theta_j} \right)\: dx,
\label{eq:A4}
\end{eqnarray}
the generalized Kullback-Leibler distance of $D(p \mid p')$ between
the two distributions $p$ and $p'$ has been introduced:
\begin{equation}
D(p \mid p') = K(p \mid p')+ K(p' \mid p),
\label{eq:A5}
\end{equation}
with
\begin{eqnarray}
K(p \mid p') &=& \int p(x)^q \:[\ln_q p(x) - \ln_q p'(x)]\: dx, 
\nonumber \\
&=& - \:\frac{1}{(q-1)} 
\left[ 1-\int p(x)^q\:p'(x)^{1-q} \:dx \right],
\label{eq:A6}
\end{eqnarray}
where $p(x)=p(x;\{\theta_i \})$ and $ \{ \theta_i \}$ denotes 
a set of parameters specifying the distribution.
In the limit of $q \rightarrow1$, $g_{ij}$ given by Eq. (\ref{eq:A4})
reduces to the conventional Fisher information matrix.
It should be remarked that Csisz\'{a}r \cite{Csiszar72}
had proposed the generalized divergence measure given by
\begin{eqnarray}
D_C(p \mid p') &=& \int 
\left[p'(x) f\left( \frac{p(x)}{p'(x)} \right)
+ p(x) f\left( \frac{p'(x)}{p(x)} \right) \right] dx,
\label{eq:A7}
\end{eqnarray}
where $f(x)$ is assumed to be a convex function with the condition $f(1)=0$.
For $f(p)=p \ln p$, Eq. (\ref{eq:A7}) yields the 
conventional Kullback-Leibler divergence
\cite{Kullback59} given by
\begin{eqnarray}
D_{KL}(p \mid p') &=& \int 
\left[p(x) \ln\left( \frac{p(x)}{p'(x)} \right)
+ p'(x) \ln\left( \frac{p'(x)}{p(x)} \right) \right] dx.
\label{eq:A8}
\end{eqnarray}
Equation (\ref{eq:A7}) for $f(p)=(q-1)^{-1}(p^q-p)$ leads to 
the generalized Kullback-Leibler distance given by Eqs. (\ref{eq:A5}) 
and (\ref{eq:A6}).
The generalized divergence given by Eq. (\ref{eq:A6}), 
which is in conformity with the Tsallis entropy,
is equivalent to the $\alpha$-divergence of Amari \cite{Amari00}
with $q=(1-\alpha)/2$ \cite{HHasegawa,Ohara07}.
The escort probability and the generalized Fisher information matrix
are discussed in Refs. \cite{Abe03,Naudts04}.
The Fisher information entropy in the Cram\'{e}r-Rao inequality has been 
studied for nonextensive systems \cite{Pennini98,Naudts04,Pennini04}.

Extensive studies on the Tsallis and Fisher entropies
have been made for reaction-diffusion systems, by using the MEM 
with exact stationary and dynamical solutions for nonlinear FPE
\cite{Plastino95,Tsallis95,Borland99,Plastino00}.
These studies nicely unify the concept of normal, super- 
and sub-diffusions by a single picture.

The purpose of the present paper is to investigate the
stationary and dynamical properties of the information entropies 
in the coupled Langevin model which
has been widely adopted for a study of various stochastic systems 
(for a recent review, see \cite{Lindner04}).
The Langevin model subjected to multiplicative noise is known to be 
one of typical nonextensive systems \cite{Tsallis04}. 
Recently the coupled Langevin model subjected to additive
and multiplicative noise has been discussed 
with the use of the augmented moment method \cite{Hasegawa07b}
which is the second-moment method for local and global variables
\cite{Hasegawa03,Hasegawa07a}.
We will obtain the probability distribution of the nonextensive, 
coupled Langevin model by using the Fokker-Planck equation (FPE) method
with the mean-field approximation.
We have made a detailed study on effects on the stationary 
information entropies of additive and multiplicative white noise,
external force, input signal, couplings and the number of constituent 
elements in the adopted model. By solving the FPE both by the proposed
analytical scheme and by the partial difference equation (PDE) method,
we have investigated the transient responses to an input signal 
and an external force which are applied to the stationary state. 

The outline of the paper is as follows.
In Sec. 2, we describe the adopted, $N$-unit coupled Langevin model.
%subjected to additive and multiplicative white noise,
%for which the probability distribution is given
%by the FPE with the mean-field approximation.
Analytical expressions for the Tsallis entropy and
generalized Fisher information entropy in some limiting cases are presented.
Numerical model calculations of stationary and dynamical entropies 
are reported. In Sec. 3, discussions are presented on
the entropy flux and entropy production and
on a comparison between $q$-moment and normal-moment methods
in which averages are taken over the escort and normal distributions, 
respectively. 
Section 4 is devoted to our conclusion.
In the Appendix, we summarize the information entropies calculated
with the use of the probability distribution derived by the MEM.
The Cram\'{e}r-Rao inequality in nonextensive systems is shown to
be expressed by the {\it extended} Fisher information entropy 
which is different from the {\it generalized} Fisher entropy.
We will discuss effects of additive and multiplicative {\it colored} noise
on information entropies, by using the result recently
obtained by the functional-integral method \cite{Hasegawa08}.

\section{Coupled Langevin model}
\subsection{Adopted model}

We have adopted the $N$-unit coupled Langevin model subjected to
additive and multiplicative white noise given by
\begin{eqnarray}
\frac{dx_i}{dt}\!\!&=&\!\!F(x_i) + \beta \xi_i(t)
+ \alpha G(x_i) \eta_i(t)+I_i(t),
\label{eq:F1} 
\end{eqnarray}
with 
\begin{equation}
I_i(t) = \frac{J}{(N-1)}\sum_{j(\neq i)}[x_j(t)-x_i(t)]+I(t).
\hspace{1cm}\mbox{($i=1$ to $N$)}
\label{eq:F2}
\end{equation}
Here $F(x)$ and $G(x)$ denote arbitrary functions of $x$,
$J$ the coupling strength, $I(t)$ an external input,
$\alpha$ and $\beta$ are the strengths of multiplicative
and additive noise, respectively, and
$\eta_i(t)$ and $\xi_i(t)$ express zero-mean Gaussian white
noises with correlations given by
\begin{eqnarray}
\langle \eta_i(t)\:\eta_j(t') \rangle 
&=& \delta_{ij} \delta(t-t'),\\
\langle \xi_i(t)\:\xi_j(t') \rangle 
&=& \delta_{ij} \delta(t-t'),\\
\langle \eta_i(t)\:\xi_j(t') \rangle &=& 0.
\end{eqnarray}

We have adopted the mean-field approximation for $I_i(t)$ given by
\begin{eqnarray}
I_i(t) &\simeq& \hat{J} [\mu_q(t)-x_i(t)]+I(t),
\label{eq:F3}
\end{eqnarray}
with 
\begin{eqnarray}
\hat{J} &=&  \frac{J N}{(N-1)},\\
\mu_q(t) &=& \frac{1}{N}\sum_i \:E_q[x_i(t)],
\label{eq:F4}
\end{eqnarray}
where the $E_q[\cdot]$ expresses the average over the escort
distribution to be shown below [Eqs. (\ref{eq:G3})-(\ref{eq:G4})].

\subsection{Fokker-Planck equation} 

Owing to the adopted mean-field approximation given by  Eq. (\ref{eq:F3}),
each element of the ensemble is ostensibly independent.
The total probability distribution of $p(\{ x_k \},t)$ is given by
the product of that of each element:
\begin{equation}
p(\{ x_k \},t) = \Pi_i\:p_i(x_i, t),
\label{eq:G1}
\end{equation}
where the FPE for $p_i(x_i,t)$ in the Stratonovich representation
is given by 
\begin{eqnarray}
\frac{\partial}{\partial t}\: p_i(x_i,t) 
&=&-\frac{\partial}{\partial x_i} [F(x_i)+I_i(t)] p_i(x_i,t) 
+ \left(\frac{\beta^2}{2}\right) 
\frac{\partial^2}{\partial x_i^2}p_i(x_i,t)
\nonumber \\
&+& \left( \frac{\alpha^2}{2} \right)
\frac{\partial }{\partial x_i}
G(x_i) \frac{\partial }{\partial x_i} G(x_i) p_i(x_i,t).
\label{eq:G2}
\end{eqnarray}
The expectation value of
$\mu_q(t)$ is given by
\begin{eqnarray}
\mu_q(t) &=& E_q[x_i(t)],
\end{eqnarray} 
with
\begin{eqnarray}
E_q[ x_i(t)^m  ] &=& \int P_{iq}(x_i,t) \: x_i^m \:dx_i, 
\hspace{0.5cm}\mbox{($m=1,2$)}
\label{eq:G3}
\end{eqnarray}
where the escort probability distribution $P_{iq}(x_i,t)$
is given by
\begin{eqnarray}
P_{iq}(x_i,t) &=& \frac{1}{c_{iq}(t)}\:p_i(x_i,t)^q, 
\label{eq:F9}\\
c_{iq}(t) &=& \int \:p_i(x_i,t)^q\:dx_i.
\label{eq:G4}
\end{eqnarray}
It is noted that $\mu_q(t)$ and $p_i(x_i,t)$ are self-consistently
determined from Eqs. (\ref{eq:F3}) and (\ref{eq:G2}).
The relevant fluctuation (variance) of $\sigma_q(t)^2$ is given by
\begin{equation}
\sigma_q(t)^2 
= E_q[(x_i-\mu_q)^2].
%= \int P_{iq}(x_i,t) \: (x_i-\mu_q)^2 \:dx_i.
\label{eq:G5}
\end{equation}

When we adopt $F(x)$ and $G(x)$ given by
\begin{eqnarray}
F(x) &=& - \lambda \:x, 
\label{eq:G6} \\
G(x) &=& x, 
\label{eq:G7}
\end{eqnarray}
where $\lambda$ denotes the relaxation rate, the FPE for $p(x,t)$ 
is expressed by (the subscript $i$ is hereafter neglected)
\begin{eqnarray}
\frac{\partial}{\partial t}\: p(x,t) 
&=&\left(\lambda+\hat{J} + \frac{\alpha^2}{2}\right) p(x,t) 
+ \left[\left( \lambda+\hat{J} 
+ \frac{3\alpha^2}{2}\right) x -u(t) \right]
\: \frac{\partial}{\partial x} p(x,t)\nonumber \\ 
&+& \left( \frac{\alpha^2}{2} x^2+ \frac{\beta^2}{2} \right)
\frac{\partial^2}{\partial x^2}p(x,t),
\label{eq:G8}
\end{eqnarray}
with
\begin{equation}
u(t)=\hat{J} \mu_q(t)+I(t).
\label{eq:G9}
\end{equation}
From the FPE given by Eq. (\ref{eq:G8}), the stationary distribution 
is given by \cite{Hasegawa07b,Sakaguchi01,Anten02}
\begin{eqnarray}
\ln p(x)&\propto& 
- \left(\frac{2 \lambda+ 2 \hat{J} +\alpha^2}{2 \alpha^2}\right)
\ln (\alpha^2 x^2 +\beta^2)
+ Y(x), \\
&\propto& -\left( \frac{1}{q-1}\right) 
\ln \left[1+(q-1)\left( \frac{x^2}{2 \phi^2} \right) \right]
+ Y(x),
\label{eq:G10} 
\end{eqnarray}
with
\begin{eqnarray}
q &=& 1 +\frac{2 \alpha^2}{(2 \lambda+ 2\hat{J}+ \alpha^2)}, 
\label{eq:G11}\\
\phi^2 &=& \frac{\beta^2}{(2 \lambda+ 2\hat{J} +\alpha^2)}, 
\label{eq:G12}\\
Y(x)&=& \left(\frac{2u}{\alpha \beta} \right) 
\tan^{-1}\left(\frac{\alpha x}{\beta} \right), \\
u &=& \hat{J} \mu_q+I, 
\label{eq:G13}
\end{eqnarray} 
where the entropic index is given for $1 \leq q < 3$.
Equation (\ref{eq:G10}) yields the $q$-Gaussian distribution given by
\begin{eqnarray}
p(x) &=& \frac{1}{Z_q} 
\exp_q\left(-\frac{x^2}{2 \phi^2} \right) \: e^{Y(x)},
\label{eq:G14} 
\end{eqnarray}
with
\begin{eqnarray}
Z_q &=& \int 
\exp_q\left(-\frac{x^2}{2 \phi^2} \right) \: e^{Y(x)} \: dx,
\label{eq:G15}
\end{eqnarray}
where $\exp_q(x)$ stands for the $q$-exponential function defined by
$\exp_q(x) \equiv [1+(1-q)x]_+^{1/(1-q)}$ where
$[y]_+=y$ for $y \geq 0$ and 0 for $y < 0$.

Some limiting cases of Eqs. (\ref{eq:G14}) are examined in the following.

\noindent
(1) For $\alpha=0$ and $\beta \neq 0$ 
({\it i.e.} additive noise only)
\begin{eqnarray}
p(x) &=& \frac{1}{\sqrt{2 \pi \sigma_1^2}} 
\: e^{-(1/2 \sigma_1^2)(x-\mu_1)^2}, 
\label{eq:G16}\
%\\
%Z_1 &=& \sqrt{2 \pi} \sigma,
\label{eq:G17} 
\end{eqnarray}
%with which Eqs. (\ref{eq:G3})  and (\ref{eq:G5}) yield
which yield
\begin{eqnarray}
\mu_1 &=& \frac{2 \phi^2 u}{\beta^2}= \frac{u}{(\lambda+\hat{J})}, \\
\sigma_1^2 &=& \phi^2 = \frac{\beta^2}{2(\lambda+\hat{J})}.
\label{eq:G18}
\end{eqnarray}

\noindent
(2) For $\alpha \neq 0$, $\beta = 0$,  
({\it i.e.} multiplicative noise only)
\cite{Hasegawa07b,Sakaguchi01,Anten02},
\begin{eqnarray}
p(x) &=& \frac{1}{Z_q} \mid x \mid^{-\delta}
e^{-\kappa/x} \: \Theta\left( x/\kappa \right),
\label{eq:G19} 
\hspace{1cm} \mbox{for $u \neq 0$} \\
&\propto& \mid x \mid^{-\delta},
\hspace{3cm} \mbox{for $u = 0$}
\end{eqnarray}
with
\begin{eqnarray}
Z_q &=& \frac{\Gamma(\delta-1)}{\kappa^{\delta-1}}
= \frac{\Gamma((3-q)/(q-1))}{\kappa^{(3-q)/(q-1)}},
\hspace{1cm} \mbox{for $u \neq 0$}, 
\label{eq:G24} \\
\delta &=& \frac{2}{(q-1)},
%=\frac{(2 \lambda+2\hat{J}+\alpha^2)}{\alpha^2}, 
\label{eq:G20}\\
\kappa &=& \frac{2u}{\alpha^2}=\frac{2(\hat{J}\mu_q+I)}{\alpha^2},
\label{eq:G21}
\end{eqnarray}
where $\Gamma(x)$ and $\Theta(x)$ denotes the gamma and
Heaviside functions, respectively, and 
%$\Theta(x)=1$ for $x > 0$ and zero otherwise;
$Z_q$ diverges for $u=0$. 
For $u \neq 0$, Eqs. (\ref{eq:G19}) and (\ref{eq:G24}) yield
\begin{eqnarray}
\mu_q &=& \frac{\: q(q-1)}{2} \kappa, \\
\sigma_q^2 &=& \frac{q^2 (q-1)^3}{4(3-q)} \kappa^2.
\label{eq:G26}
\end{eqnarray}
The distribution given by Eq. (\ref{eq:G19}) has a peak at
$x=\kappa/\delta=\mu_q/q$.

\noindent
(3) For $\alpha \neq 0$, $\beta \neq 0$, $u=\hat{J}\mu_q+I= 0$ 
({\it i.e.} without coupling and external input), 
\cite{Hasegawa07b,Sakaguchi01,Anten02}
\begin{eqnarray}
p(x) &=& \frac{1}{Z_q} 
\exp_q\left(-\frac{x^2}{2 \phi^2} \right),
\label{eq:G22} \\
Z_q &=& \left(\frac{2 \phi^2}{q-1}\right)^{1/2}
B\left(\frac{1}{2},\frac{1}{q-1}-\frac{1}{2} \right),
\label{eq:G23}
\end{eqnarray}
%Equations (\ref{eq:G22}) and (\ref{eq:G23}) lead to
which lead to
\begin{eqnarray}
\mu_q&=&0, \\
\sigma_q^2 &=& \frac{2 \phi^2}{(3-q)}
=\frac{\beta^2}{2 \lambda}.
\label{eq:G27}
\end{eqnarray}
It is noted that when we adopt normal moments
averaged over the $q$-Gaussian given by
\begin{eqnarray}
E[x(t)^m] &=& \int p(x,t) \: x(t)^m \:dx,
\label{eq:G25}
\end{eqnarray}
in stead of the $q$-moments given by Eq. (\ref{eq:G3}),
its stationary variance is given by 
$\sigma^2=E[(x-E[x])^2]=\beta^2/2(\lambda-\alpha^2)$
which diverges at $\lambda=\alpha^2$ \cite{Hasegawa07b}.

\subsection{Tsallis entropy} 

With the use of the total distribution of $p(\{ x_i\})$ given 
by Eq. (\ref{eq:G1}), the Tsallis entropies of single-unit 
and $N$-unit ensembles are given by
\begin{eqnarray}
S_q^{(1)} &=& \left( \frac{1-c_q}{q-1} \right), 
\label{eq:X2} \\
S_q^{(N)} &=& \left( \frac{1-\Pi_i \:c_{iq}}{q-1} \right)
= \left( \frac{1-c_q^N}{q-1} \right),
\label{eq:X3}
\end{eqnarray}
with
\begin{eqnarray}
c_{q} &=& c_{iq} = \int p_i(x_i)^q \:dx_i.
\label{eq:X4}
\end{eqnarray}
Eliminating $c_q$ from Eqs. (\ref{eq:X2}) and (\ref{eq:X3}), we get
\begin{eqnarray}
S_q^{(N)} &=& \sum_{k=1}^N C_k^N 
(-1)^{k-1} (q-1)^{k-1} (S_q^{(1)})^k, 
\label{eq:X5} 
\\
&=& N S_q^{(1)}-\frac{N(N-1)}{2}(q-1)(S_q^{(1)})^2+ \cdot\cdot, 
\label{eq:X6}
\end{eqnarray}
where $C_k^N=N!/(N-k)!\:k!$. Equation (\ref{eq:X5}) shows
that the Tsallis entropy is non-extensive except for $q=1.0$, 
for which $S_q^{(N)}$ reduces to the extensive Boltzmann-Gibbs-Shannon 
entropy: $S_1^{(N)}=N\:S_1^{(1)}$.

Substituting the stationary distributions given by Eqs. (\ref{eq:G16}),
(\ref{eq:G19}) and (\ref{eq:G22}) to Eq. (\ref{eq:A1}), 
we get the analytic expression for
the Tsallis entropy of a single unit given by
\begin{eqnarray}
S_q^{(1)} &=& \left( \frac{1}{2} \right) [1 + \ln (2 \pi \sigma_q^2)],
\hspace{1cm}\mbox{for $\alpha=0$, $\beta \neq 0$} 
\label{eq:H1}\\
&=& \left( \frac{1-c_q}{q-1} \right), 
\hspace{0.5cm}\mbox{for $\alpha \neq 0$}
\label{eq:H2}
\end{eqnarray}
with
\begin{eqnarray}
c_q &=& \frac{1}{Z_q^q} \frac{\Gamma(q \delta-1)}{(q\kappa)^{q\delta-1}}
= \frac{1}{Z_q^q}\frac{\Gamma((q+1)/(q-1))}{(q \kappa)^{(q+1)/(q-1)}},
\hspace{0.5cm}\mbox{for $\alpha \neq 0$, $\beta=0$, $u \neq 0$} 
\label{eq:H4} \\
&=& \frac{1}{Z_q^q}\left( \frac{2 \phi^2}{q-1} \right)^{1/2}
B\left( \frac{1}{2}, \frac{q}{q-1}-\frac{1}{2} \right)
= \left( \frac{3-q}{2} \right)\:Z_q^{1-q}, \nonumber \\
&&\hspace{5cm}\mbox{for $\alpha \neq 0$, $\beta \neq 0$, $u=0$}
%= \nu \:Z_q^{1-q},
\label{eq:H3}
\end{eqnarray}
where $B(a,b)$ stands for the beta function,
and $Z_q$ in Eqs. (\ref{eq:H4}) and (\ref{eq:H3}) 
are given by Eqs. (\ref{eq:G24}) and (\ref{eq:G23}), respectively. 

\subsection{Generalized Fisher information entropy} 

We consider the generalized Fisher information entropy given by 
\begin{eqnarray}
g_q &=&
q \int p(x) \left(\frac{\partial \ln p(x)}{\partial \theta} \right)^2 \: dx.
\label{eq:H7}
\end{eqnarray}
%with $\theta=\mu_q$.
From Eqs. (\ref{eq:G1}) and (\ref{eq:H7}), the generalized 
Fisher entropy for the $N$-unit system is given by
\begin{eqnarray}
g_q^{(N)} &=& q\: E\left[\left(\frac{\partial \ln p(x)}
{\partial \theta}\right)^2 \right], 
\label{eq:X7} \\
&=& q \int \cdot\cdot \int  
\left(\sum_i \frac{\partial  \ln p_i(x_i)}{\partial \theta} \right)^2 
\:\Pi_i\: p_i(x_i)dx_i, 
\label{eq:X8} \\
&=& q \sum_i \int 
\left( \frac{\partial \ln p_i(x_i)}{\partial \theta}\right)^2 
\:p_i(x_i)\:dx_i
+ \Delta g_q, 
\label{eq:X9} \\
&=& N g_q^{(1)}, 
\label{eq:X10}
\end{eqnarray}
because the cross term $\Delta g_q$ of Eq. (\ref{eq:X9}) vanishes:
\begin{eqnarray}
\Delta g_q 
&=& q \sum_{i(\neq j)} \sum_j \;\int 
\left( \frac{\partial \ln p_i(x_i)}{\partial \theta}\right) p_i(x_i) \;dx_i
\; \int 
\left( \frac{\partial \ln  p_j(x_j)}{\partial \theta}\right)p_j(x_j) \;dx_j,  
\label{eq:X11} \\
&=& q \sum_{i(\neq j)} \sum_j 
\;\int \frac{\partial p_i(x_i)}{\partial \theta}\;dx_i
\; \int \frac{\partial p_j(x_j)}{\partial \theta}\;dx_j,  
\label{eq:X12} \\
&=& q \sum_{i(\neq j)} \sum_j
\;\frac{\partial}{\partial \theta} \int p_i(x_i) \:dx_i 
\; \frac{\partial}{\partial \theta} \int p_j(x_j) \:dx_j, 
\label{eq:X13} \\
&=& 0, \label{eq:X14}
\end{eqnarray}
where $g_q^{(1)}$ stands for the generalized Fisher entropy in a single 
subsystem. The generalized Fisher information entropy is 
extensive in the nonextensive system as shown in \cite{HHasegawa}:
\begin{equation}
g_q^{(N)}=N g_q^{(1)}.
\label{eq:H10}
\end{equation}

The probability distribution
$p(x)$ obtained by the FPE for our Langevin model
is determined by the six parameters of $\lambda$, $\alpha$, 
$\beta$, $J$, $I$ and $N$.
When adopt $\theta=I$ in Eq. (\ref{eq:H7}), for example, we get
the generalized Fisher entropy given by
\begin{eqnarray}
g_q & = & 
q\left( E\left[\left( \frac{\partial Y(x)}{\partial I} \right)^2 \right]
-E\left[ \left( \frac{\partial Y(x)}{\partial I }\right) \right]^2 \right),
\end{eqnarray}
where $E[ \cdot]$ expresses the average over $p(x)$ [Eq. (\ref{eq:G25})].

Alternatively we have adopted the generalized Fisher entropy given by 
\begin{eqnarray}
g_q &=&
q \int p(x) \left(\frac{\partial \ln p(x)}{\partial x} \right)^2 \: dx,
\label{eq:H11}
\end{eqnarray}
which is obtainable for $g_q$ with $\theta=\mu_q$ in Eq. (\ref{eq:H7}) 
if $p(x)$ is given by the MEM [Eqs. (\ref{eq:B6}) and (\ref{eq:D3})].
Although $p(x)$ derived by the FPE is 
not explicitly specified by $\mu_q$ and $\sigma_q^2$,
we have employed Eq. (\ref{eq:H11}) in our following discussion, 
expecting it is meaningful for both cases of the FPE and MEM.
Substituting the stationary distributions given by Eqs. (\ref{eq:G16}),
(\ref{eq:G19}) and (\ref{eq:G22}) to Eq. (\ref{eq:H11}), 
we get the analytic expression for the generalized Fisher
entropy for $N=1$ given by 
\begin{eqnarray}
g_q^{(1)} &=& \left( \frac{1}{\sigma_1^2}\right)
= \frac{2(\lambda+\hat{J})}{\beta^2},
\hspace{2.5cm}\mbox{for $\alpha=0$, $\beta \neq 0$} 
\label{eq:H8} \\
&=& \left( \frac{q}{\kappa^2} \right)\delta (\delta-1)(\delta+2)
%= \frac{4 q^2 (3-q)}{(q-1)^3 \kappa^2}
= \frac{q^4}{\sigma_q^2}, 
\hspace{0.5cm}\mbox{for $\alpha \neq 0$, $\beta=0$, $u \neq 0$}
\label{eq:H12} \\
&=& \left( \frac{2 q}{(q-1)\phi^2} \right)
\frac{B(\frac{3}{2},\frac{1}{q-1}+\frac{1}{2})}
{B(\frac{1}{2},\frac{1}{q-1}-\frac{1}{2})}
= \frac{1}{\sigma_q^2},
%= \frac{2 \lambda}{\beta^2}.
\hspace{0.5cm}\mbox{for $\alpha \neq 0$, $\beta \neq 0$, $u = 0$}
\label{eq:H9}
\end{eqnarray}
where $\sigma_q^2$ in Eqs. (\ref{eq:H12}) and (\ref{eq:H9}) are given by 
Eqs. (\ref{eq:G26}) and (\ref{eq:G27}), respectively.

\subsection{Stationary properties}

\subsubsection{Calculation method}

The adopted Langevin model includes six parameters of $\lambda$,
$\alpha$, $\beta$, $J$, $I$ and $N$.
The dependence of the Tsallis entropy and generalized Fisher information 
entropy on these parameters have been studied by numerical methods.
We have calculated the distribution $p(x)$ by the FPE
[Eqs. (\ref{eq:G14}) and (\ref{eq:G15})], 
and also by direct simulations (DSs) for the Langevin
model [Eqs. (\ref{eq:F1}) and (\ref{eq:F2})] 
with the Heun method: DS results are averages of 100 trials. 

\subsubsection{Model calculations}

Figures \ref{figA}(a)-\ref{figA}(c) show three examples of the stationary
distribution $p(x)$ for $(I,J)=$ (0.0, 0.0), (0.0, 0.5) and (0.5, 0.5)
with $\lambda=1.0$, $\alpha=0.5$, $\beta=0.5$ and $N=100$.
Solid curves show the results calculated with the use of
the FPE whereas dashed curves those of DSs for the Langevin equation:
both results are in good agreement and indistinguishable.
When the coupling strength is increased from $J=0.0$
to $J=0.5$ with $I=0.0$, the width of $p(x)$ is decreased because
of a decreased $\phi^2$ in Eq. (\ref{eq:G12}).
When an input of $I=0.5$ is applied, $p(x)$ changes its position
by an amount of about 0.5 with a slight variation of its shape:
$p(x)$ for $(I,J)=(0.5,0.5)$ is not a simple translational shift
of $p(x)$ for $(I,J)=(0.0,0.5)$.

In the following, we will discuss model calculations of the dependence 
on $\alpha$, $\beta$, $I$, $J$ and $N$, 
whose results are shown in Figs. \ref{figB}, \ref{figC}, \ref{figD}, 
\ref{figE} and \ref{figF}, respectively
(dotted curves in the frames (a) and (b) in Figs. \ref{figB}-\ref{figE}
will be explained in Sec. 3.6.1).

\vspace{0.5cm}
\noindent
{\bf $\alpha$ dependence}

First we show $\mu_q$, $\sigma_q^2$,
$S_q$ and $g_q$ in Figs. \ref{figB}(a), \ref{figB}(b), \ref{figB}(c) 
and \ref{figB}(d), respectively, plotted as a function of $\alpha^2$
for $I=0.0$ (chain curves), $I=0.5$ (dashed curves) and $I=1.0$ (solid curves)
with $\lambda=1.0$, $\beta=0.5$ and $J=0.0$.
Figure \ref{figB}(a) shows that the $\alpha$ dependence of $\mu_q$ is very weak.
We note in Fig. \ref{figB}(b) that for $I=0.5$ and $I=1.0$,
$\sigma_q^2$ is linearly increased with increasing $\alpha^2$
though $\sigma_q^2$ is independent of $\alpha$ for $I=0.0$.
Figure \ref{figB}(c) shows that with increasing $\alpha^2$,
$S_q$ is increased with broad maxima at $\alpha^2 \sim 0.8$ for
$I=1.0$ and at $\alpha^2 \sim 1.5$ for $I=0.5$.
With increasing $\alpha^2$ from $\alpha^2=0$, 
in contrast, $g_q$ is decreased for $I=0.5$ and $I=1.0$ 
with broad minima, whereas $g_q$ is independent of $\alpha^2$ for $I=0.0$.
For larger $I$, $S_q$ and $g_q$ have stronger $\alpha^2$ dependence. 

\vspace{0.5cm}
\noindent
{\bf $\beta$ dependence}

Figures \ref{figC}(a), \ref{figC}(b), \ref{figC}(c) 
and \ref{figC}(d) show $\mu_q$, $\sigma_q^2$,
$S_q$ and $g_q$, respectively, plotted as a function of $\beta^2$ 
for $I=0.0$ (chain curves), $I=0.5$ (dashed curves) and $I=1.0$ (solid curves)
with $\lambda=1.0$, $\alpha=0.5$ and $J=0.0$.
With increasing $\beta$, $\mu_q$ has no changes although
$\sigma_q^2$ is linearly increased.
With increasing $\beta^2$ from $\beta^2=0.0$, 
$S_q$ ($g_q$) is significantly increased (decreased).
This trend is more significant for $I=0.0$ than for $I=0.5$ and $I=1.0$.

\vspace{0.5cm}
\noindent
{\bf $I$ dependence}

The $I$ dependence of $\mu_q$, $\sigma_q^2$, $S_q$ and $g_q$ 
are shown in Figs. \ref{figD}(a)-\ref{figD}(b) 
for $\alpha=0.0$ (chain curves), $\alpha=0.5$ (dashed curves)
and $\alpha=1.0$ (solid curves) with $\lambda=1.0$. 
The gradient of $\mu_q$ versus $I$ is slightly larger for larger $\alpha$.
In the case of $\alpha=0.0$, $\sigma_q^2$, $S_q$ and $g_q$ 
are independent of $I$ [see Eqs. (\ref{eq:H1}) and (\ref{eq:H8})].
With increasing $I$ for finite $\alpha$, $S_q$ is increased
while $g_q$ is decreased. 

\vspace{0.5cm}
\noindent
{\bf $J$ dependence}

We show the $J$ dependence of $\mu_q$, $\sigma_q^2$, $S_q$ and $g_q$ 
in Figs. \ref{figE}(a)-\ref{figE}(b), for $I=0.0$ (chain curves), 
$I=0.5$ (dashed curves) and $I=1.0$ (solid curves) with
$\lambda=1.0$, $\alpha=0.5$, $\beta=0.5$ and $N=100$.
We note that $\mu_q$ is independent of $J$.
With increasing $J$, $\sigma_q^2$ and $S_q$ are linearly decreased
whereas $g_q$ is increased.

\vspace{0.5cm}
\noindent
{\bf $N$ dependence}

%and $S_q^{(1)}$ denotes the entropy of a single element ($N=1$).
Figure \ref{figF} shows the Tsallis entropy per element, $S_q^{(N)}/N$,
given by [Eq. (\ref{eq:X6})]
\begin{eqnarray}
\frac{S_q^{(N)}}{N}
%&=& \sum_{k=1}^N \left(\frac{(N-1)!}{(N-k)!\:k!} \right)
%(-1)^{k-1} (q-1)^{k-1} (S_q^{(1)})^k, \\
&=& S_q^{(1)}-\frac{1}{2}(N-1)(q-1)(S_q^{(1)})^2+ \cdot\cdot, 
\end{eqnarray}
for $\alpha=0.0$ (dotted curve), $\alpha=0.1$ (solid curve), 
$\alpha=0.5$ (dashed curve) and $\alpha=1.0$ (chain curve) 
with $\lambda=1.0$, $\beta=0.5$, $I=0.0$ and $J=0.0$.
Note that for $\alpha=0.0$ ($q=1.0$), the system is extensive
because $S_1^{(N)}/N=S_1^{(1)}$.
For finite $\alpha$, however, it is nonextensive: 
$S_q^{(N)}/N$ is more significantly decreased for larger $\alpha$,
though the generalized Fisher information entropy $g_q$ is extensive 
[Eq. (\ref{eq:H10})].

\subsection{Dynamical properties}
\subsubsection{Analytical method for the FPE}

In order to discuss the dynamical properties of the
entropies, we have to calculate the 
time-dependent probability $p(x,t)$, solving the FPE given 
by Eq. (\ref{eq:G8}). In the case of $q=1.0$, we may obtain
the exact solution of the Gaussian distribution given by 
\begin{equation}
p(x,t)= \frac{1}{\sqrt{2 \pi \:\sigma_1(t)^2}}
\;e^{-[x-\mu_1(t)]^2/2 \sigma_1(t)^2},
\label{eq:K1}
\end{equation}
where $\mu_1(t)$ and $\sigma_1(t)^2$ satisfy equations of motion given by
\begin{eqnarray}
\frac{d \mu_1(t)}{dt} &=& -\lambda \mu_1(t)+ I, 
\label{eq:K2} \\
\frac{d \sigma_1(t)^2}{dt} &=& -2 (\lambda+\hat{J}) \sigma_1(t)^2 + \beta^2.
\label{eq:K3}
\end{eqnarray}

In order to obtain an analytical solution of the FPE for $q > 1.0$, 
we have adopted the following method:

\noindent
(1) Starting from an equation of motion for $n$th $q$-moment of $E_q[x^n]$
given by
\begin{eqnarray}
\frac{d E_q[x^n]}{d t} &=& \frac{d}{d t}
\int P_q(x,t)\:x^n \:dx,\\
&=& \frac{q}{c_q}\int 
\left( \frac{\partial p(x,t)}{\partial t}\right)\:p(x,t)^{q-1}\:x^n \:dx
-\frac{1}{c_q}\left( \frac{d c_q}{dt} \right) E_q[x^n],
\label{eq:L2} \\
\frac{d c_q}{dt} 
&=& q \int \left( \frac{\partial p(x,t)}{\partial t}\right)\:p(x,t)^{q-1}\:dx,
\label{eq:L0}
\end{eqnarray} 
%After some manipulations with substituting 
%$ \partial p(x,t)/\partial t$ in the FPE
%to Eqs. (\ref{eq:L2}) and (\ref{eq:L0}), 
we have obtained equations of motion for $\mu_q(t)$ ($=E_q[x]$) 
and $\sigma_q(t)^2$ ($=E_q[x^2]-E[x]^2$), % with $J=0$, 
valid for $O(\alpha^2)$ and $O(\beta^2)$, as given by \cite{Hasegawa07b}
\begin{eqnarray}
\frac{d \mu_q(t)}{dt} &\simeq& -\lambda \mu_q(t) + I, 
\label{eq:L3} \\
\frac{d \sigma_q(t)^2}{dt} &\simeq& -2 (\lambda+\hat{J}) \sigma_q(t)^2 
+ \alpha^2 \mu_q(t)^2+\beta^2.
\label{eq:L4}
\end{eqnarray}
Equations (\ref{eq:L3}) and (\ref{eq:L4}) lead to the stationary solution 
given by
\begin{eqnarray}
\mu_q &=& \frac{I}{\lambda}, 
\label{eq:L5} \\
\sigma_q^2 &=& \frac{(\alpha^2 \mu_q^2 +\beta^2)}{2 (\lambda+\hat{J})}
=  \frac{(\alpha^2 I^2/\lambda^2 +\beta^2)}{2 (\lambda+\hat{J})}.
\label{eq:L6}
\end{eqnarray}

\noindent
(2) We rewrite the distribution of $p(x)$ given by 
Eqs. (\ref{eq:G11})-(\ref{eq:G15}) in terms of $\mu_q$, $\sigma_q^2$ and $q$, as
\begin{eqnarray}
p(x) &=& \frac{1}{Z_q} 
\left[1-(1-q)\left( \frac{x^2}{2 \phi^2}\right) \right]^{\frac{1}{1-q}}
%\exp_q\left(-\frac{x^2}{2 \phi^2} \right)
e^{Y(x)}, 
\label{eq:L7} 
\end{eqnarray}
with
\begin{eqnarray}
Y(x) &=& \left( \frac{(3-q)\: \mu_q} 
{\sqrt{2(q-1) \phi^2}} \right)
\tan^{-1} \left(\sqrt{\frac{(q-1)}{2 \phi^2}}x \right), 
\label{eq:L8} \\
\phi^2 &=& \left( \frac{3-q}{2}\right) \sigma_q^2 
-\left( \frac{q-1}{2}\right) \mu_q^2, 
\label{eq:L9}
%
%\nu &=& \frac{3-q}{2}.
%\label{eq:L10}
\end{eqnarray}
where $Z_q$ expresses the normalization factor [Eq. (\ref{eq:G15})].
In deriving Eqs. (\ref{eq:L7})-(\ref{eq:L9}), we have employed
relations given by
\begin{eqnarray}
\alpha^2 &=& \frac{2(q-1)(\lambda + \hat{J})}{(3-q)},
\label{L15} \\
\beta^2 &=& 2(\lambda+\hat{J})
\left[\sigma_q^2-\frac{(q-1)\mu_q^2}{(3-q)} \right],
\end{eqnarray}
which are obtained from Eqs. (\ref{eq:G11}), (\ref{eq:L5}) and (\ref{eq:L6}).

\noindent
(3) Then we have assumed that a solution of $p(x,t)$ of the FPE given 
by Eq. (\ref{eq:G8}) is expressed by Eqs. (\ref{eq:L7})-(\ref{eq:L9}) 
in which stationary $\mu_q$ and $\sigma_q^2$ are replaced by time-dependent 
$\mu_q(t)$ and $\sigma_q(t)^2$ with equations of motion given 
by Eqs. (\ref{eq:L3}) and (\ref{eq:L4}).

Dotted curves in the frames (a) and (b) of Figs. 3-6 express the 
results of stationary $\mu_q$ and $\sigma_q^2$ calculated by 
Eqs. (\ref{eq:L5}) and (\ref{eq:L6}) for some typical sets of parameters. 
They are in good agreement with those shown by solid curves obtained 
with the use of the stationary distribution of $p(x)$ given by Eq. (\ref{eq:G14}).

As will be shown shortly, the approximate, analytical method given 
by Eqs. (\ref{eq:L3}), (\ref{eq:L4}), (\ref{eq:L7})-(\ref{eq:L9}) 
provides fairly good results for dynamics of $\mu_q(t)$, $\sigma_q(t)^2$ 
and $S_q(t)$, and also for that of $g_q(t)$ except for the transient period.

\subsubsection{Partial difference equation method}

In order to examine the validity of the analytical method discussed above, 
we have adopted also the numerical method, using the partial 
difference equation (PDE) derived from Eq. (\ref{eq:G8}), as given by
\begin{eqnarray}
p(x,t+b) &=& p(x,t)
+\left(\lambda+\hat{J}+ \frac{\alpha^2}{2} \right) b \:p(x,t)
\nonumber \\
&+&\left[x \left(\lambda+\hat{J} 
+ \frac{3 \alpha^2}{2} \right)-u(t) \right]
\left(\frac{b}{2 a}\right)[p(x+a)-p(x-a)] \nonumber \\
&+& \left(\frac{\alpha^2}{2} x^2+ \frac{\beta^2}{2} \right)
\left(\frac{b}{a^2}\right)[p(x+a,t)+p(x-a,t)-2p(x,t)],
\label{eq:K4}
\end{eqnarray}
with
\begin{equation}
u(t) = \hat{J}\:\mu_q(t)+I(t),
\label{eq:K5}
\end{equation}
where $a$ and $b$ denote incremental steps of $x$ and $t$, respectively.

We impose the boundary condition:
\begin{eqnarray}
p(x,t)=0, \hspace{1cm}\mbox{for $ \mid x \mid \ge x_m$}
\label{eq:K6}
\end{eqnarray}
with $x_m=5$, and the initial condition of $p(x,0)=p_0(x)$ where $p_0(x)$ 
is the stationary distribution given by Eqs. (\ref{eq:G14}) and (\ref{eq:G15}).
We have chosen parameters of $a=0.05$ and $b=0.0001$ such as to satisfy 
the condition: $(\alpha^2 x_m^2 b/2 a^2) < 1/2$, which is required for
stable, convergent solutions of the PDE.

\subsubsection{Model calculations}

\noindent
{\bf Response to $I(t)$}

We apply the pulse input signal given by
\begin{equation}
I(t) = \Delta I \:\Theta(t-2)\Theta(6-t),
\label{eq:K7}
\end{equation}
where $\Delta I=1.0$ and $\Theta(t)$ denotes the Heaviside 
function: $\Theta(t)=1$ for $t > 0$ and zero otherwise.
Figure \ref{figG} shows the time-dependent distribution at various 
$t$ for $\lambda=1.0$, $\alpha=0.5$, $\beta=0.5$ and $J=0.0$.
Solid and dashed curves express the results of the PDE method
and the analytical method (Sec. 3.6.1), respectively.
%by Eqs. (\ref{eq:L3}), (\ref{eq:L4}) and (\ref{eq:L7})-(\ref{eq:L9}), 
When input of $\Delta I$ is applied at $t=2.0$, 
the distribution is gradually changed, moving rightward.
The results of the analytical method are in good 
agreement with those obtained by the PDE method,
except for $t = 3$ and $t = 7$.

This change in $p(x,t)$ induces changes in $\mu_q(t)$, $\sigma_q(t)^2$, 
$S_q(t)$ and $g_q(t)$, whose time dependences are shown 
in Figs. \ref{figH}(a) and \ref{figH}(b), 
solid and dashed curves expressing the results of 
the PDE method and the analytical method, respectively.
By an applied pulse input, $\mu_q$, $\sigma_q^2$ and $S_q$ are increased 
while $g_q$ is decreased.
The result for $S_q(t)$ of the analytical method is in fairly good 
agreement with that obtained by the PDE method. 
The calculated $g_q(t)$ of the analytical method
is also in good agreement with that of the PDE method
besides near the transient periods at $t \simg 2$ and $t \simg 6$
just after the input signal is on and off.
This is expected due to the fact that $g_q(t)$ given by Eq. (\ref{eq:H11})
is sensitive 
to a detailed form of $p(x,t)$ because it is expressed by an integration
of $(\partial p(x,t)/\partial x)^2$ over $p(x,t)$, 
while $S_q(t)$ is obtained by a simple integration of $p(x,t)^q$.

For a comparison, we show by chain curves, the results of the PDE method
when the step input given by
\begin{equation}
I(t) = \Delta I \:\Theta(t-2),
\label{eq:K8}
\end{equation}
is applied.
The relaxation time of $S_q$ and $g_q$ is about 2.0.

It is noted that input signal for $\alpha=0$ induces 
no changes in $S_q(t)$ and $g_q(t)$, which has been already realized
in the stationary state as shown by 
chain curves in Figs. \ref{figD}(c) and \ref{figD}(d). 

\vspace{0.5cm}
\noindent
{\bf Response to $\lambda(t)$}

We modify the relaxation rate as given by 
\begin{equation}
\lambda = 1.0 + \Delta \lambda \: \Theta(t-2)\Theta(6-t),
\label{eq:K9}
\end{equation}
which expresses an application of an external force 
of $\Delta F$ ($=-\Delta \lambda \:x$) at $2 \leq t < 6$
with $\Delta \lambda=0.5$.
Figure \ref{figI}(a) and \ref{figI}(b) show 
the time dependence of $\sigma_q^2$, $S_q$ and $g_q$ 
with $\alpha=0.5$, $\beta=0.5$, $I=0.0$ and $J=0.0$
for which $\mu_q=0$.
Solid and dashed curves express the results of the PDE method
and the analytical method, respectively.
When an external force is applied, $\sigma_q^2$ and
$S_q$ are decreased whereas $g_q$ is increased.
The results of the analytical method are in good agreement
with those of the PDE method.
The relaxation times of $S_q$ and $g_q$ are 0.47 and 0.53,
respectively.

\section{Discussion}

\subsection{Maximum-entropy method}

In the preceding Sec. 2, we have discussed the information entropies
by using the probability distribution obtained by the FPE for the Langevin model.
It is worthwhile to compare it with the probability
distribution derived by the MEM.
The variational condition for the Tsallis entropy given by Eq. (\ref{eq:A1}) 
is taken into account with the three constraints: 
a normalization condition and $q$-moments of $x$ and $x^2$, as given by 
\cite{Tsallis95,Borland99,Plastino00,Hasegawa06}
\begin{eqnarray}
1 &=& \int p(x)\: dx, 
\label{eq:B1}
\\
\mu_q &=& E_q[x]=\int P_q(x) \:x \: dx, 
\label{eq:B2}
\\
\sigma_q^2 &=& E_q[(x-\mu_q)^2]
=\int P_q(x) \:(x-\mu_q)^2\:dx,
\label{eq:B3}
\end{eqnarray}
where $E_q[\cdot]$ expresses the average over the escort probability 
of $P_q(x)$ given by
\begin{eqnarray}
P_q(x) &=& \frac{p(x)^q}{c_q},
\label{eq:B4} 
\\
c_q &=& \int p(x)^q \: dx,
\label{eq:B5}
\end{eqnarray}
the entropic index $q$ being assumed to be $0 < q < 3$.
After some manipulations, we get the $q$-Gaussian (non-Gaussian)
distribution given by
\cite{Hasegawa06}
\begin{eqnarray}
p(x) &=& \frac{1}{Z_q} 
\exp_q\left(-\frac{(x-\mu_q)^2}{2 \nu \sigma_q^2} \right),
\label{eq:B6}
\end{eqnarray}
with
\begin{eqnarray}
\nu &=& \left(\frac{3-q}{2}\right),
\label{eq:B7} \\
Z_q &=& \int 
\exp_q\left(-\frac{(x-\mu_q)^2}{2 \nu \sigma_q^2} \right) \:dx, \\
&=& \left(\frac{2 \nu \sigma_q^2}{q-1} \right)^{1/2}
B\left(\frac{1}{2}, \frac{1}{q-1}-\frac{1}{2} \right),
\hspace{1cm}\mbox{for $1< q < 3$}
\label{eq:B8} 
\\
&=& \sqrt{2 \pi} \sigma_1,
\label{eq:B9}
\hspace{5.5cm}\mbox{for $q=1$} 
\\
&=& \left(\frac{2 \nu \sigma_q^2}{1-q} \right)^{1/2}
B\left(\frac{1}{2}, \frac{1}{1-q}+1 \right).
\hspace{1cm}\mbox{for $0< q < 1$}
\label{eq:B10}
\end{eqnarray}
%where $B(a,b)$ denotes the beta function.
In the limit of $q \rightarrow 1$, 
$p(x)$ in Eq. (\ref{eq:B6}) becomes the Gaussian distribution:
\begin{equation}
p(x) = \frac{1}{\sqrt{2 \pi} \sigma_1} e^{-(x-\mu_1)^2/2 \sigma_1^2}.
\label{eq:B11}
\end{equation}
The probability distribution given by Eq. (\ref{eq:B6}) 
derived from the MEM is
different from that of Eq. (\ref{eq:G14}) obtained by the FPE
although both expressions are equivalent for $\mu_q=I=J=0$
with $\nu \sigma_q^2=\phi^2$. 
Note that the former is defined 
for $0 < q < 3$ while the latter is valid for $1 \leq q < 3$.

A comparison between the probability distributions
obtained by the FPE and MEM is made in Figs. \ref{figR} and \ref{figS}.
Figure \ref{figR}(a) shows the probability distributions
calculated by the FPE of the Langevin model
for $\alpha=0.0$,
0.5, 1.0, 1.5 and 2.0, which yield 
$(q,\sigma_q^2)=(1.0, 0.125)$,
$(1.222, 0.25)$, $(1.667, 0.625)$, $(2.059, 1.25)$ 
and $(2.333, 2.125)$, respectively, with $\mu_q=1.0$ for $I=1.0$,
$\lambda=1.0$, $\beta=0.5$ and $J=0.0$
[Eqs. (\ref{eq:L5}) and (\ref{eq:L6})].  
Figure \ref{figR}(b) shows corresponding distributions calculated by the
MEM with the respective parameters of $q$, $\mu_q$ and $\sigma_q$.
For $\alpha=0.0$ ($q=1.0$), both distributions of the FPE
and MEM are Gaussian centered at $x=\mu_q=1.0$.
For $q\neq 1.0$, $p(x)$ of the FPE becomes asymmetric
with respect to $x=\mu_q$ while that of the MEM is still symmetric.
The peak position of $p(x)$ of the MEM is at $x=\mu_q=1.0$ independent of
$q$ while that of the FPE moves leftward with increasing $\alpha$.
It is noted that $p(x)$ of the FPE for $\alpha \neq 0$ and $\beta=0$ 
given by Eq. (\ref{eq:G19}) has a peak at $x=\mu_q/q$.

Figure \ref{figS}(a) shows $p(x)$ of the FPE
for various inputs of $I=0.0$,
0.5, 1.0, 1.5 and 2.0, which yield 
$(\mu_q,\sigma_q^2)=(0.0, 0.125)$,
$(0.5, 0.25)$, $(1.0, 0.625)$, $(1.5, 1.25)$ 
and $(2.0, 2.125)$, respectively,
for $\lambda=1.0$, $\alpha=1.0$, $\beta=0.5$ and $J=0.0$
[Eqs. (\ref{eq:L5}) and (\ref{eq:L6})];
corresponding $p(x)$ of the MEM with the respective parameters of 
$q\;(=1.667)$, $\mu_q$ and $\sigma_q$ are plotted in Fig. \ref{figS}(b).
For $\mu_q=0.0$, both the distributions agree.
Although  centers of both distributions move rightward
with increasing $\mu_q$, their profiles and 
peak positions are different between the two distributions.
We note that the magnitude of $p(x)$ at $x < 0.0$ 
of the FPE is smaller than that of the MEM for $\mu_q \neq 0.0$.

The information entropies calculated with the use of the distribution
given by Eq. (\ref{eq:B6}) are summarized in the Appendix.
One of the advantages of the MEM is that its distribution is explicitly 
specified by the parameters of $(\theta_1, \theta_2)$=$(\mu_q, \sigma_q^2)$
while that of the FPE is given in an implicit way 
[{\it cf.} Eqs. (\ref{eq:L7})-(\ref{eq:L9})].
We may discuss the upper bound of estimation errors
by the Cram\'{e}r-Rao inequality, which is shown to be expressed
by the {\it extended} Fisher entropy [Eq. (\ref{eq:E0})]
but not by the generalized Fisher entropy [Eq. (\ref{eq:D0})].

In order to discuss the dynamics within the MEM for $q \neq 1.0$, 
we have once tried to obtain an analytic solution
of its distribution $p(x,t)$, assuming that it is
given by Eq. (\ref{eq:B6}):
\begin{eqnarray}
p(x,t) &=& \left(\frac{A_q}{\sqrt{\sigma_q(t)^2}} \right)
%\frac{1}{Z_q(t)} 
\exp_q\left[-\frac{(x-\mu_q(t))^2}{2 \nu \sigma_q(t)^2} \right],
\label{eq:L1}
\end{eqnarray}
where the $q$-dependent coefficient $A_q$ is determined from 
Eqs. (\ref{eq:B8})-(\ref{eq:B10}), and equations of motion for
$\mu_q(t)$ and $\sigma_q(t)^2$ are derived so as to meet the FPE
after Refs. \cite{Borland99,Plastino00}.
Unfortunately, we could not uniquely determined them:
we got two equations for $d \mu_q(t)/dt$ and three equations
for $d \sigma_q(t)^2/dt$ which are mutually not consistent
(except for $q=1.0$).
This implies that the exact analytic solution of the FPE
is not given by Eq. (\ref{eq:L1}).
Indeed, the exact solution for $\beta=J=0$ in Eq. (\ref{eq:G8})
does not have a functional form given by Eq. (\ref{eq:L1}) \cite{Fa03}. 

\subsection{Entropy flux and entropy production}

It is interesting to discuss the entropy flux and entropy
production from the time derivative of the Tsallis entropy given by
\begin{eqnarray}
\frac{d S_q(t)}{dt} &=& - \left( \frac{q}{q-1} \right)
\int p(x,t)^{q-1} \left( \frac{\partial p(x,t)}{\partial t} \right)\: dx, 
\label{eq:P1} \\
&=& Q_F+Q_A+Q_M,
\label{eq:P2}
\end{eqnarray}
with
\begin{eqnarray}
Q_F &=& q \int p(x,t)^q \left( \frac{d F(x)}{d x} \right) \:dx
+ q(q-1) \int p(x,t)^{q} 
\left( \frac{\partial \ln p(x,t)}{\partial x}\right) F(x)  \: dx, 
\label{eq:P3} \\
Q_A &=& \left( \frac{\alpha^2 q}{2} \right) \int p(x,t)^q 
\left( \frac{\partial \ln p(x,t)}{\partial x} \right)^2 \: dx, 
\label{eq:P4} \\
Q_M &=& \left( \frac{\beta^2}{2}\right) \int p(x,t)^q 
\left[ q \left( \frac{\partial \ln p(x,t)}{\partial x} \right)^2 G(x)^2
- \left( \frac{d G(x)}{d x} \right)^2 
- \frac{d^2G(x)}{dx^2} G(x) \right] \: dx. \nonumber 
\label{eq:P5}\\
&&
\end{eqnarray}
Here $Q_F$ denotes the entropy flux, and $Q_A$ and $Q_M$ stand for 
entropy productions due to additive and multiplicative noise, respectively. 

By using the stationary distribution given by Eq. (\ref{eq:G22}),
we get $Q_F$, $Q_A$ and $Q_M$ in the stationary state 
with $I=J=0$ ({\it i.e.} without couplings and external input):
\begin{eqnarray}
Q_F &=& - \frac{\lambda q}{Z_q^q} 
\left( \frac{2 \sigma_q^2}{q-1} \right)^{1/2}
B\left(\frac{1}{2}, \frac{1}{q-1}+\frac{1}{2} \right) \nonumber \\
&+& \frac{\lambda q(q-1)}{\sigma_q^2 Z_q^q}
\left(\frac{2 \sigma_q^2}{q-1} \right)^{3/2}
B\left(\frac{3}{2}, \frac{1}{q-1}+\frac{1}{2} \right),
\label{eq:P6} \\
Q_A &=& \frac{\beta^2 \:q}{2 \sigma_q^4 Z_q^q}
\left(\frac{2 \sigma_q^2}{q-1} \right)^{3/2}
B\left(\frac{3}{2}, \frac{1}{q-1}+\frac{3}{2} \right), 
\label{eq:P7}\\
Q_M &=& \frac{\alpha^2 \: q}{2 \sigma_q^4 Z_q^q}
\left(\frac{2 \sigma_q^2}{q-1} \right)^{5/2}
B\left(\frac{5}{2}, \frac{1}{q-1}+\frac{1}{2} \right) \nonumber \\
&-& \frac{\alpha^2}{2 Z_q^q}
\left(\frac{2 \sigma_q^2}{q-1} \right)^{1/2}
B\left(\frac{1}{2}, \frac{1}{q-1}+\frac{1}{2} \right),
\label{eq:P8}
\end{eqnarray}
where $Z_q$ is given by Eq. (\ref{eq:G23}).
Equations (\ref{eq:P6})-(\ref{eq:P8}) satisfy the stationary condition: 
$Q_F+Q_A+Q_M=0$.

It is worthwhile to examine the limit of 
$ \alpha \rightarrow 0$ ($q \rightarrow 1.0$),
in which Eqs. (\ref{eq:P1}), (\ref{eq:P6})-(\ref{eq:P8}) yield
\begin{eqnarray}
\frac{d S_1(t)}{d t} &=& - \int \frac{\partial p(x,t)}{\partial t} 
\ln p(x,t) \:dx, 
\label{eq:P9} \\
&=& Q_F+Q_A,
\label{eq:P10}
\end{eqnarray}
with
\begin{eqnarray}
Q_F &=& \int p(x,t) \left( \frac{d F(x)}{dx} \right) \: dx,
\label{eq:P11} \\
Q_A &=& \left( \frac{\alpha^2}{2} \right) \int p(x,t) 
\left( \frac{\partial \ln p(x,t)}{\partial x} \right)^2 \:dx.
\label{eq:P12}
\end{eqnarray}
With noticing the relation:
$ \lim_{\mid z \mid \rightarrow \infty} 
[\Gamma(z+a)/\Gamma(z) z^a]=1$ \cite{Abramo}, 
we may see that Eqs. (\ref{eq:P11}) and (\ref{eq:P12}) lead to
$Q_F= -Q_A=-\lambda$ and $d S_1/d t=0$ in the limit of $q \rightarrow 1$. 

In the opposite limit of $\beta \rightarrow 0$,
Eqs. (\ref{eq:P6})-(\ref{eq:P8}) yields  that each of $Q_F$. $Q_A$ and $Q_M$ is 
proportional to $1/\beta^{(q-1)}$ and then divergent in this limit,
though $Q_F+Q_A+Q_M =0$.
It is noted that $Q_A=\lambda$ for $\alpha \rightarrow 0$ and 
$\beta \rightarrow 0$ \cite{Daems99,Bag01,Bag02}.
 
We present some model calculations of $Q_F$, $Q_A$ and $Q_M$ 
in the stationary state, which are shown in Fig. \ref{figL} 
as a function of $\alpha$ for $\beta=0.1$ (dashed curves),
$\beta=0.5$ (chain curves) and $\beta=1.0$ (solid curves).
We note that $Q_F < 0$ and $Q_A+Q_M > 0$.
With increasing $\alpha$, $Q_F$ is decreased 
in the case of $\beta=0.1$, while it is increased
in the cases of $\beta=0.5$ and 1.0.
Bag \cite{Bag02} showed that $Q_F$ is always decreased with increasing
$\alpha$ which disagrees with our result mentioned above:
Eqs. (\ref{eq:P6})-(\ref{eq:P8}) are rather different 
from Eqs. (36) and (37) in Ref. \cite{Bag02} where
non-Gaussian properties of the distribution is not properly
taken into account.

\subsection{$q$-moment and normal-moment methods}

In Refs. \cite{Hasegawa07b,Hasegawa07a}, we have discussed equations 
of motion for normal moments of $\mu$ ($=E[x]$) and
$\sigma^2$ ($=E[(x-\mu)^2]$) [Eq. (\ref{eq:G25})]
in the Langevin model with $J=0$, as given by
\begin{eqnarray}
\frac{d \mu(t)}{dt} 
&=& -\left(\lambda- \frac{\alpha^2}{2} \right)\mu(t)+ I, 
\label{eq:L11} \\
\frac{d \sigma(t)^2}{dt} &=& -2 (\lambda-\alpha^2) \sigma(t)^2 
+ \alpha^2 \mu(t)^2+\beta^2.
\label{eq:L12}
\end{eqnarray}
These equations of motion are rather different from 
those for the $q$-moments of $\mu_q$ and $\sigma_q^2$
given by Eqs. (\ref{eq:L3}) and (\ref{eq:L4}). 
Indeed, Eqs. (\ref{eq:L11}) and (\ref{eq:L12}) 
yield stationary normal moments given by 
\begin{eqnarray}
\mu &=& \frac{I}{(\lambda-\alpha^2/2)}, 
\label{eq:L13} \\
\sigma^2 &=& \frac{(\alpha^2 \mu^2+\beta^2)}{2(\lambda-\alpha^2)},
\label{eq:L14}
\end{eqnarray}
which are different from the stationary $q$-moments of $\mu_q$ and $\sigma_q^2$  
given by Eqs. (\ref{eq:L5}) and (\ref{eq:L6}),
and which diverge at $\lambda=\alpha^2/2$ and $\lambda=\alpha^2$,
respectively.

The time dependence of $\mu(t)$ and $\sigma(t)^2$ becomes considerably 
different from those of $\mu_q(t)$ and $\sigma_q(t)^2$ 
for an appreciable value of $\alpha$. 
Figure \ref{figP}(a), (b), (c) and (d) show some examples
of $\mu_q(t)$, $\sigma_q(t)^2$, $\mu(t)$ and $\sigma(t)^2$, respectively,
when a pulse input given by Eq. (\ref{eq:K7}) is applied
with $\alpha=0.2$ (chain curves), $\alpha=0.5$ (dashed curves)
and $\alpha=0.8$ (solid curves).
Although $\mu_q(t)$ is independent of $\alpha$, $\sigma_q(t)^2$, $\mu(t)$
and $\sigma(t)^2$ are much increased at $2 \leq t < 6$ 
for larger $\alpha$. In particular, $\sigma(t)^2$ is significantly 
increased because of the $\alpha^2$ term in Eq. (\ref{eq:L12}).

\subsection{Effects of colored noise}

We have so far considered additive and multiplicative 
{\it white} noise.
In our recent paper \cite{Hasegawa08}, 
we have taken into account the effect of {\it colored} noise
by employing the functional-integral method.
We have assumed
%{\it colored} noise, applying
%the functional-integral method to  
the Langevin model subjected to
additive ($\chi$) and multiplicative ($\phi$) colored noise 
given by 
\begin{eqnarray}
\frac{dx(t)}{dt}\!\!&=&\!\! -\lambda x(t) + \chi(t)
+ x(t) \phi(t) + I(t), 
\label{eq:R1} 
\end{eqnarray}
with
\begin{eqnarray}
\frac{d \chi(t)}{d t}
&=& - \frac{1}{\tau_a} [\chi(t)- \beta \;\xi(t)], 
\label{eq:R2} \\
\frac{d \phi(t)}{d t}
&=& -\frac{1}{\tau_m} [\phi(t)-  \alpha \;\eta(t)],
\label{eq:R3} 
\end{eqnarray}
where $\tau_a$ and $\beta$ ($\tau_m$ and $\alpha$) 
express the relaxation time and strength of additive (multiplicative)
noise, respectively, and
$\xi$ and $\eta$ stand for independent zero-mean Gaussian white noise.
By applying the functional-integral method to the Langevin model
given by Eqs. (\ref{eq:R1})-(\ref{eq:R3}), we have obtained 
the effective one-variable FPE, from which 
the effective Langevin model is derived as
\cite{Hasegawa08}
\begin{eqnarray}
\frac{dx(t)}{dt}\!\!&=&\!\! -\lambda x(t)
+ \tilde{\beta} \: \xi(t)
+ \tilde{\alpha}(t) \:x(t) \: \eta(t) + I(t),
\label{eq:R4}  
\end{eqnarray}
with
\begin{eqnarray}
\tilde{\beta}^2 &=& \frac{\beta^2}{(1+\lambda \tau_a) }, 
\label{eq:R5} \\
\tilde{\alpha}(t)^2 &=& \frac{\alpha^2}{[1+ \tau_m I(t)/\mu(t) ]}.
\label{eq:R6} 
%\hspace{1.0cm}\mbox{($i= 1, 2$)} %\nonumber
\end{eqnarray}
Here $\mu(t)$ is given by Eq. (\ref{eq:L11}) %and (\ref{eq:L12})
with $\alpha=\tilde{\alpha}$, %and $\beta=\tilde{\beta}$,
from which $\tilde{\alpha}$ is determined in a self-consistent way.

In the stationary state where $\mu=I/(\lambda - \tilde{\alpha}^2/2)$ 
given by Eq. (\ref{eq:L13}) with $\alpha = \tilde{\alpha}$, 
we get $\tilde{\alpha}$ from Eq. (\ref{eq:R6}):
\begin{eqnarray}
\tilde{\alpha}^2 &=& \frac{\alpha^2}{[1+\tau_m(\lambda-\tilde{\alpha}^2/2)]}, 
\label{eq:R7} \\
&=& \frac{1}{\tau_m}
\left[(1+\lambda \tau_m)-\sqrt{(1+\lambda \tau_m)^2-2 \tau_m \alpha^2}\right]. 
\label{eq:R8} 
\end{eqnarray}
%For $ \tau_m \alpha^2/2(1+\lambda \tau_m)^2 \ll 1$, 
We get an approximate expression given by \cite{Hasegawa08}
\begin{eqnarray}
\tilde{\alpha}^2 
& \simeq & \frac{\alpha^2}{(1+\lambda \tau_m)},
\label{eq:R9} 
%\equiv (\tilde{\alpha}^{APP})^2,
\hspace{1.0cm}\mbox{for $ \tau_m \alpha^2/2(1+\lambda \tau_m)^2 \ll 1$} 
\end{eqnarray}
which is shown to be a good approximation
both for $\tau_m \ll (1/\lambda, \: 2/\alpha^2)$ 
and $\tau_m \gg (1/\lambda, \: \alpha^2/2 \lambda^2)$ \cite{Hasegawa08}.
Equations (\ref{eq:R5}) and (\ref{eq:R9}) show that effects of additive 
and multiplicative colored noise are described by 
$\tilde{\alpha}^2$ and $\tilde{\beta}^2$ which are reduced
by factors of $(1 + \lambda \tau_a)$ and $(1 + \lambda \tau_m$),
respectively, from original values of $\alpha^2$ and $\beta^2$.

The $\tau_a$ dependence of $S_q$ and $g_q$ is plotted 
in Fig. \ref{figQ}(a) and \ref{figQ}(b) with $\tau_m=0.0$
for $I=0.0$ (chain curves), $I=0.5$ (dashed curves) and 
$I=1.0$ (solid curves) with $\lambda=1.0$, $\alpha=0.5$
and $\beta=0.5$. 
We note that with increasing $\tau_a$, $g_q$ is much increased
for smaller $I$ whereas $S_q$ is much decreased for smaller $I$.
The dependence of $S_q$ and $g_q$ on $\tau_a$ may be understood
from their $\beta$ dependence
shown in Figs. \ref{figC}(c) and \ref{figC}(d). 
The $\tau_m$ dependence of $S_q$ and $g_q$ is plotted 
in Fig. \ref{figQ}(a) and \ref{figQ}(b) with $\tau_a=0.0$
with $\lambda=1.0$, $\alpha=0.5$ and $\beta=0.5$. 
With increasing $\tau_m$,
$S_q$ ($g_q$) is decreased (increased) for $I=0.5$ and $I=1.0$
while no changes for $I=0.0$.
These behavior may be explained from 
Figs. \ref{figB}(c) and \ref{figB}(d) showing the $\alpha$ dependence of
$S_q$ and $g_q$.

\section{Conclusion}
We have discussed stationary and dynamical properties
of the Tsallis and Fisher entropies in nonextensive systems.
Our calculation for the $N$-unit coupled Langevin model subjected to
additive and multiplicative noise has shown the followings:
 
\noindent
(i) the dependence of $S_q$ and $g_q$ on the parameters of
$\lambda$, $\alpha$, $\beta$, $I$, $J$ and $N$ in the coupled 
Langevin model are clarified (Figs. \ref{figB}-\ref{figF}), and
%with increasing $J$, both $\sigma^2$ and $S_q$
%are decreased while $g_q$ is increased (Fig. \ref{figE}), 

%\noindent
%(ii) multiplicative noise significantly modifies 
%the dependence of $S_q$ and $g_q$ on $I$ (Fig. \ref{figD}), and

\noindent
(ii) dynamical properties are well described by the analytical method for the FPE 
proposed in Sec. 2.6.1, which shows that the relaxation times in transient 
responses of $S_q$ and $g_q$ to a change in $\lambda$ are short ($\tau \sim 0.5$)
while those in $I$ are fairly long ($\tau \sim 2$).

\noindent
The difference between the parameter dependence of $S_q$
and $g_q$ in the item (i) arises from the fact that
$S_q$ provides us with a global measure of ignorance
while $g_q$ a local measure of positive amount of information
\cite{Frieden98}.

We have calculated the information entropies also by using
the probability distribution derived 
by the MEM, from which we get the followings:

\noindent
(iii) $p(x)$ derived by the MEM is rather
different from that of the FPE for $\mu_q \neq 0$ 
(Figs. \ref{figR} and \ref{figS}),
for which the information entropies of the MEM are independent of $\mu_q$
while those of the FPE depend on $\mu_q$ ({\it i.e.} $I$), and

\noindent
(iv) the Cram\'{e}r-Rao inequality is expressed by
the extended Fisher entropy [Eq. (\ref{eq:E0})] which is different
from the generalized Fisher entropy [Eq. (\ref{eq:D0})]
derived from the generalized Kullback-Leibler divergence [Eq. (\ref{eq:A6})].

\noindent
The item (iv) has not been clarified in previous studies on the Fisher 
entropies in nonextensive systems %\cite{Pennini98,Naudts04,Pennini04}.
\cite{Plastino95}-\cite{Masi06}.

The Langevin model has been employed for a study
of a wide range of stochastic systems \cite{Lindner04}. 
Quite recently, the present author has proposed the
generalized rate-code model for neuronal ensembles
which is described by the coupled Langevin-type equation
\cite{Hasegawa07d,Hasegawa07c}. 
It would be interesting to discuss the dynamics of 
information entropies in such neural networks,
which is left for our future study.

\section*{Acknowledgments}
This work is partly supported by
a Grant-in-Aid for Scientific Research from the Japanese 
Ministry of Education, Culture, Sports, Science and Technology.  

\vspace{1.5cm}

\appendix

\noindent
{\large\bf Appendix: The Maximum-entropy method}
\renewcommand{\theequation}{A\arabic{equation}}
\setcounter{equation}{0}

By using the probability distribution given by Eq. (\ref{eq:B6})
derived by the MEM, we have calculated the information entropies,
which are summarized in the Appendix.

\vspace{0.5cm}
\noindent
{\bf Tsallis entropy}

With the use of Eqs. (\ref{eq:A1}) and (\ref{eq:B6}), 
the Tsallis entropy is given by
\begin{eqnarray}
S_q &=& \left( \frac{1}{2}\right) [1+ \ln(2 \pi \sigma_q^2)],
\hspace{1cm}\mbox{for $q = 1$} 
\label{eq:C1}
\\
&=&\left(\frac{1-c_q}{q-1} \right),
\hspace{2cm}\mbox{for $q \neq1$} 
\label{eq:C2}
\end{eqnarray}
with
\begin{eqnarray}
c_q &=& \frac{1}{Z_q^q}\left(\frac{2 \nu \sigma_q^2}{q-1} \right)^{1/2}
B\left(\frac{1}{2}, \frac{q}{q-1}-\frac{1}{2} \right),
\hspace{1cm}\mbox{for $1 < q < 3$} 
\label{eq:C3}
\\
&=& \frac{1}{Z_q^q} \left(\frac{2 \nu \sigma_q^2}{1-q} \right)^{1/2}
B\left(\frac{1}{2}, \frac{q}{1-q}+1 \right),
\hspace{1cm}\mbox{for $0 < q < 1$} 
\label{eq:C4}
\end{eqnarray}
which yield
\begin{equation}
c_q = \nu \: Z_q^{1-q}.
\hspace{1cm}\mbox{for $0 < q < 3$}
\end{equation}
Here $Z_q$ for $0 < q <1$ and $1 < q < 3$ 
are given by Eqs. (\ref{eq:B8}) and (\ref{eq:B10}), 
respectively.

\vspace{0.5cm}
\noindent
{\bf Generalized Fisher entropy}

The distribution $p(x)$ given by Eq. (\ref{eq:B6}) 
is characterized by
two parameters of $(\theta_1, \theta_2)=(\mu_q, \sigma_q^2)$.
By using Eqs. (\ref{eq:A4}) and (\ref{eq:B6}), we obtain the component of
the generalized Fisher information matrix {\sf G} given by
\cite{Pennini98}-\cite{Masi06}
\begin{eqnarray}
g_{ij}&=& q \:E\left[ 
\left( \frac{\partial \ln p(x)}{\partial \theta_i} \right)
\left( \frac{\partial \ln p(x)}{\partial \theta_j} \right) \right], 
\label{eq:D0} \\
&=& q\:E[ (X_i - E[X_i])( X_j - E[X_j]) ],
\label{eq:D1}
\hspace{1cm}\mbox{for $i,j=1,2$}
\end{eqnarray}
with
\begin{eqnarray}
X_i= \frac{\partial }{\partial \theta_i} 
\ln\left[\exp_q \left(-\frac{(x-\mu_q)^2}{2 \nu \sigma_q^2} \right) \right],
\label{eq:D2}
\end{eqnarray}
where $E[\cdot]$ denotes the average over the $q$-Gaussian 
distribution of $p(x)$ whereas $E_q[\cdot]$ stands for the average over 
the escort distribution of $P_q(x)$.
Substituting the probability given by Eq. (\ref{eq:B6}) 
to Eq. (\ref{eq:D0}), we get
\begin{eqnarray}
g_{11} &=& 
q \int p(x) \left(\frac{\partial \ln p(x)}{\partial \mu_q} \right)^2 \: dx, \\
&=& q \int p(x) \left(\frac{\partial \ln p(x)}{\partial x} \right)^2 \: dx, 
\label{eq:D3} \\
&=& \left(\frac{2q}{\nu \sigma_q^2 (q-1)}\right)
\frac{B(\frac{3}{2},\frac{1}{(q-1)}+\frac{1}{2})}
{B(\frac{1}{2},\frac{1}{(q-1)}-\frac{1}{2})},
\hspace{1cm}\mbox{for $1 < q < 3$}
\label{eq:D4} 
\\
&=& \frac{1}{\sigma_q^2},
\hspace{7.5cm}\mbox{for $q=1$} 
\label{eq:D5}
\\
&=& \left( \frac{2q}{\nu \sigma_q^2 (1-q)} \right)
\frac{B(\frac{3}{2},\frac{1}{(1-q)}-1)}
{B(\frac{1}{2},\frac{1}{(1-q)}+1)},
\hspace{1cm}\mbox{for $0 < q < 1$}
\label{eq:D6}
\end{eqnarray}
which yield
\begin{eqnarray}
g_{11} &=& \frac{1}{\sigma_q^2}.
\hspace{1cm}\mbox{for $0 < q < 3$}
\label{eq:D7} 
\end{eqnarray}
A similar calculation leads to the (2,2)-component given by
\begin{eqnarray}
g_{22} &=& 
q \int p(x) \left(\frac{\partial \ln p(x)}
{\partial \sigma_q^2} \right)^2 \: dx, \\
&=& \left( \frac{3-q}{4\sigma_q^4} \right).
\hspace{1cm}\mbox{for $0 < q < 3$}
\label{eq:D11} 
\end{eqnarray}

The generalized Fisher information matrix is expressed by
\[\sf {G}=\left(
\begin{array}{cc}
\frac{1}{\sigma_q^2} & 0 \\
0 & \frac{(3-q)}{4 \sigma_q^4} \\
\end{array}
\right),\]
whose inverse 
is given by
\[\sf G^{-1}=\left(
\begin{array}{cc}
\sigma_q^2 & 0\\
0 & \frac{4 \sigma_q^4}{(3-q)} \\
\end{array}
\right).\]
In the limit of $q=1$, the matrix reduces to
\[\sf G = \left(
\begin{array}{cc}
\frac{1}{\sigma_q^2} & 0 \\
0 & \frac{1}{2 \sigma_q^4} \\
\end{array}
\right).
\hspace{1cm}\mbox{for $q=1$} 
\]

\vspace{0.5cm}
\noindent
{\bf Extended Fisher entropy: Cram\'{e}r-Rao inequality}

Next we discuss the Cram\'{e}r-Rao inequality
in nonextensive systems. 
For the escort distribution given by Eq. (\ref{eq:B4})
which satisfies Eqs. (\ref{eq:B2}) and (\ref{eq:B3}) with 
\begin{equation}
1 = E_q[1] = \int P_q(x)\: dx,
\end{equation}
we get
the Cram\'{e}r-Rao inequality 
\cite{Frieden98,Pennini98,Naudts04,Pennini04}
\begin{equation}
\sf V \geq \sf \tilde{G}^{-1}.
\label{eq:E1}
\end{equation}
Here $\sf V$ denotes the covariance error matrix whose
explicit expression will be given shortly, and
{\sf \^{G}} is referred to as the {\it extended} Fisher 
information matrix whose
components are expressed by
\begin{eqnarray}
\tilde{g}_{ij} &=& E_q \left[
\left(\frac{\partial \ln P_q(x)}{\partial \theta_i}\right) 
\left(\frac{\partial \ln P_q(x)}{\partial \theta_j}\right) 
\right], 
\hspace{1cm}\mbox{for $i,j=1,2$} \label{eq:E0}\\
&=& E_q \left[ (\tilde{X}_i-E_q[\tilde{X}_i])
(\tilde{X}_j-E_q[\tilde{X}_j]) \right],
\end{eqnarray}
with
\begin{eqnarray}
\tilde{X}_i &=& \frac{\partial}{\partial \theta_i} [q \: \ln p(x)], \\
&=& q (X_i-E[X_i]),
\label{eq;E2}
\end{eqnarray}
$X_i$ being given by Eq. (\ref{eq:D2}).
Note that $\tilde{g}_{ij}$ % in Eq. (\ref{eq:E6}) 
is different from $g_{ij}$ given by Eq. (\ref{eq:D0})
except for $q=1.0$.
The (1,1) component of {\sf \^{G}} is given by
\begin{eqnarray}
\tilde{g}_{11} &=& E_q \left[\left(\frac{\partial \ln P_q(x)}
{\partial \mu_q}\right)^2 \right], 
\label{eq:E3} 
\\
&=&  \left( \frac{q^2}{c_q} \right)  
\int p(x)^q \left( \frac{\partial \ln p(x)}
{\partial x}\right)^2 \:dx,
\label{eq:E4} \\
&=& \left(\frac{2q^2}{\nu \sigma_q^2 (q-1)}\right)
\frac{B(\frac{3}{2},\frac{q}{(q-1)}+\frac{1}{2})}
{B(\frac{1}{2},\frac{q}{(q-1)}-\frac{1}{2})},
\hspace{1cm}\mbox{for $1 < q < 3$} 
\label{eq:E5}
\\
&=& \frac{1}{\sigma_q^2},
\hspace{7cm}\mbox{for $q=1$} 
\label{eq:E6}
\\
&=& \left( \frac{2q^2}{\nu \sigma_q^2 (1-q)} \right)
\frac{B(\frac{3}{2},\frac{q}{(1-q)}-1)}
{B(\frac{1}{2},\frac{q}{(1-q)}+1)},
\hspace{1cm}\mbox{for $1/2 < q < 1$}
\label{eq:E7}
\end{eqnarray}
which lead to
\begin{eqnarray}
\tilde{g}_{11} &=& \frac{q(q+1)}{(3-q)(2q-1)\sigma_q^2}.
\hspace{1cm}\mbox{for $1/2 < q < 3$}
\label{eq:E8}
\end{eqnarray}
Similarly, the (2,2) component of $\sf \tilde{G}$ is given by
\begin{eqnarray}
\tilde{g}_{22} &=& E_q \left[\left(\frac{\partial \ln P_q(x)}
{\partial \sigma_q^2}\right)^2 \right], 
\label{eq:E10} \\
&=& \frac{(q+1)}{4 (2 q-1)\sigma_q^4}.
\hspace{1cm}\mbox{for $1/2 < q < 3$}
\label{eq:E9}
\end{eqnarray}

The extended Fisher information matrix $\sf \tilde{G}$ is expressed by
\[
\sf \tilde{G}=\left(
\begin{array}{cc}
\frac{q(q+1)}{(3-q)(2 q-1)\sigma_q^2} & 0\\
0 & \frac{(q+1)}{4(2q-1)\sigma_q^4}  \\
\end{array}
\right),
\]
whose inverse is given by
\[\sf \tilde{G}^{-1}=\left(
\begin{array}{cc}
\frac{(3-q)(2 q-1)\sigma_q^2}{q(q+1)}  & 0\\
0 & \frac{4(2q-1)\sigma_q^4}{(q+1)}\\
\end{array}
\right).
\]
A calculation of the $(i,j)$ component ($v_{ij}$) of
the covariance error matrix $\sf V$ leads to
\[{\sf V}= \left(
\begin{array}{cc}
\sigma_q^2 & 0 \\
0 & \frac{4 \sigma_q^4}{(5-3 q)} \\
\end{array}
\right).
\]
In the limit of $q=1$, the matrices reduce to
\[\sf \tilde{G}^{-1} =\sf G^{-1} = \left(
\begin{array}{cc}
\sigma_q^2 & 0 \\
0 & 2 \sigma_q^4 \\
\end{array}
\right),
\hspace{1cm}\mbox{for $q=1$} 
\]

\[{\sf V}= \left(
\begin{array}{cc}
\sigma_q^2 & 0 \\
0 & 2 \sigma_q^4 \\
\end{array}
\right).
\hspace{2cm}\mbox{for $q=1$} 
\]

Chain and solid curves in Fig. \ref{figN}(a)
express the $q$ dependence of 
$v_{11}/\sigma_q^2$ and $1/\tilde{g}_{11}\sigma_q^2$, respectively.
When $q$ is further from unity, 
$1/\tilde{g}_{11}$ is much decreased and it vanishes at 
$q=1/2$ and 3. 
The lower bond of $v_{11}$ is expressed by the
Cram\'{e}r-Rao relation because it is satisfied by $\tilde{g}_{11}$:
\begin{eqnarray}
v_{11} &=& \frac{1}{g_{11}} \geq \frac{1}{\tilde{g}_{11}}.
\hspace{1cm} \mbox{for $1/2 < q < 3$} 
\label{eq:E11}
\end{eqnarray}
Chain, dashed and solid curves in  Fig. \ref{figN}(b)
show $v_{22}/\sigma_q^4$, $1/g_{22}\sigma_q^4$ 
and $1/\tilde{g}_{22}\sigma_q^4$, respectively.
It is noted that $v_{22}$ diverges at $q=5/3$.
The following relations hold:
\begin{eqnarray}
\frac{1}{g_{22}} & > & v_{22} > \frac{1}{\tilde{g}_{22}},
\hspace{1cm} \mbox{for $1/2 < q < 1$} 
\label{eq:E12} \\
v_{22} & \geq & \frac{1}{\tilde{g}_{22}} \geq \frac{1}{g_{22}}. 
\hspace{1cm} \mbox{for $1 \leq q < 5/3$}
\label{eq:E13}
\end{eqnarray}
Equation (\ref{eq:E12}) means that $1/g_{22}$ cannot provide the lower
bound of $v_{22}$. 
Equations (\ref{eq:E11})-(\ref{eq:E13}) clearly show 
that the lower bound of $\sf V$
is expressed by the extended Fisher information matrix $\sf \tilde{G}$, 
but not by the generalized Fisher information matrix $\sf G$.

%\end{references}

\newpage

\begin{figure}
\caption{
(Color online)
Stationary distribution $p(x)$ for $(I, J)=(0.0,0.0)$,
$(0.0,0.5)$ and $(0.5, 0.5)$ with $\lambda=1.0$, 
$\alpha=0.5$ and $\beta=0.5$, calculated by the FPE 
[Eqs. (\ref{eq:G14}) and (\ref{eq:G15})] (solid curves)
and by direct simulations (DSs) for the coupled Langevin model
[Eqs. (\ref{eq:F1}) and (\ref{eq:F2})] (dashed curves).
}
\label{figA}
\end{figure}

\begin{figure}
\caption{
(Color online)
The $\alpha^2$ dependence of 
(a) $\mu_q$, (b) $\sigma_q^2$, (c) $S_q$ and (d) $g_q$
for $I=0.0$ (chain curves), $I=0.5$ (dashed curves)
and $I=1.0$ (solid curves) with $\lambda=1.0$, $\beta=0.5$ and $J=0.0$.
Dotted curves in (a) and (b) express the analytical result
given by Eqs. (\ref{eq:L5}) and (\ref{eq:L6}) for $I=1.0$.
}
\label{figB}
\end{figure}

\begin{figure}
\caption{
(Color online)
The $\beta^2$ dependence of 
(a) $\mu_q$, (b) $\sigma_q^2$, (c) $S_q$ and (d) $g_q$
for $I=0.0$ (chain curves), $I=0.5$ (dashed curves)
and $I=1.0$ (solid curves) with $\lambda=1.0$, $\alpha=0.5$ and $J=0.0$.
Dotted curves in (a) and (b) express the analytical result
given by Eqs. (\ref{eq:L5}) and (\ref{eq:L6}) for $I=1.0$.
}
\label{figC}
\end{figure}

\begin{figure}
\caption{
(Color online)
The $I$ dependence of 
(a) $\mu_q$, (b) $\sigma_q^2$, (c) $S_q$ and (d) $g_q$
for $\alpha=0.0$ (chain curves), $\alpha=0.5$ (dashed curves)
and $\alpha=1.0$ (solid curves) with $\lambda=1.0$, $\beta=0.5$ and $J=0.0$.
Dotted curves in (a) and (b) express the analytical result
given by Eqs. (\ref{eq:L5}) and (\ref{eq:L6}) for $\alpha=1.0$.
}
\label{figD}
\end{figure}

\begin{figure}
\caption{
(Color online)
The $J$ dependence of 
(a) $\mu_q$, (b) $\sigma_q^2$, (c) $S_q$ and (d) $g_q$
for $I=0.0$ (chain curves), $I=0.5$ (dashed curves)
and $I=1.0$ (solid curves) with $\lambda=1.0$, $\alpha=0.5$, $\beta=0.5$
and $N=100$.
Dotted curves in (a) and (b) express the analytical result
given by Eqs. (\ref{eq:L5}) and (\ref{eq:L6}) for $I=1.0$.
}
\label{figE}
\end{figure}

\begin{figure}
\caption{
%(Color online)
The $N$ dependence of the Tsallis entropy per element,
$S_q^{(N)}/N$, for $\alpha=0.0$ (dotted curve), $\alpha=0.1$ (solid curve),
$\alpha=0.5$ (dashed curve) and $\alpha=1.0$ (chain curve)
with $\lambda=1.0$, $\beta=0.5$, $I=0.0$ and $J=0.0$.
}
\label{figF}
\end{figure}

\begin{figure}
\caption{
(Color online)
The time-dependent probability distribution
$p(x,t)$ when an input pulse given by
$I(t) = \Delta I \:\Theta(t-2)\Theta(6-t)$ 
%Eq. (\ref{eq:K7}) 
is applied with
$\Delta I =1.0$, $\lambda=1.0$, $\alpha=0.5$ and $\beta=0.5$:
solid curves express the results obtained by the PDE method
and chain curves denote those 
by the analytical method described in Sec. 2.6.1.
% given by 
%Eqs. (\ref{eq:L3}), (\ref{eq:L4}), (\ref{eq:L7})-(\ref{eq:L9}).
Curves are consecutively shifted downward
by 0.25 for a clarity of the figure.
}
\label{figG}
\end{figure}

\begin{figure}
\caption{
(Color online)
The time dependence of (a) $\mu_q(t)$ and $\sigma_q(t)^2$ and
(b) $S_q(t)$ and $g_q(t)$ for an input of
$I(t) = \Delta I \:\Theta(t-2)\Theta(6-t)$ with
$\Delta I=1.0$, $\lambda=1.0$, $\alpha=0.5$, $\beta=0.5$ and $J=0.0$.
Solid curves denote the results obtained by the PDE method
and dashed curves those obtained 
by the analytical method described in Sec. 2.6.1.
%by the analytic solution
%given by Eqs. (\ref{eq:L3}), (\ref{eq:L4}), (\ref{eq:L7})-(\ref{eq:L9}).
Chain curves denote the results of the PDE method for an input
signal given by $I(t)=\Delta I \Theta(t-2)$,
results of $g_q$ and $\mu_q$ being divided by a factor of ten.
}
\label{figH}
\end{figure}

\begin{figure}
\caption{
(Color online)
The time dependence of (a) $\sigma_q(t)^2$ and (b) $S_q(t)$ and $g_q(t)$
for $\lambda(t)=1.0 + \Delta \lambda \: \Theta(t-2)\Theta(6-t)$: 
$\Delta \lambda=0.5$, $\alpha=0.5$, $\beta=0.5$, $I=0.0$ and $J=0.0$.
Solid curves denote the results obtained by the PDE method
and dashed curves those obtained 
by the analytical method described in Sec. 2.6.1,
%by the analytic solution
%given by Eqs. (\ref{eq:L3}), (\ref{eq:L4}), (\ref{eq:L7})-(\ref{eq:L9}),
results of $g_q$ being divided by a factor of ten.
}
\label{figI}
\end{figure}

\begin{figure}
\caption{
Probability distributions $p(x)$ calculated by 
(a) the FPE and (b) MEM for
$(\alpha,q,\sigma_q^2)=(0.0, 1.000, 0.125)$,
$(0.5, 1.222, 0.25)$, $(1.0, 1.667, 0.625)$, $(1.5, 2.059 1.25)$ 
and $(2.0, 2.333, 2.125)$
with $I=1.0$, $\lambda=1.0$, $\beta=0.5$, and $J=0.0$,
figures in parentheses of (a) denoting $\alpha$ values.
}
\label{figR}
\end{figure}

\begin{figure}
\caption{
Probability distributions $p(x)$ calculated by 
(a) the FPE and (b) MEM for
%$I=0.0$, 0.5, 1.0, 1.5 and 2.0, which lead to
$(\mu_q,\sigma_q^2)=(0.0, 0.125)$,
$(0.5, 0.25)$, $(1.0, 0.625)$, $(1.5, 1.25)$ 
and $(2.0, 2.125)$, respectively,
with $\lambda=1.0$, $\alpha=1.0$, $\beta=0.5$ and $J=0.0$.  
}
\label{figS}
\end{figure}

\begin{figure}
\caption{
(Color online)
The $\alpha$ dependence of 
entropy flux ($Q_F$), and entropy productions by
additive noise ($Q_A$) and multiplicative noise ($Q_M$)
for $\beta=0.1$ (dashed curves), $\beta=0.5$ (chain curves)
and $\beta=1.0$ (solid curves) with $\lambda=1.0$, 
$I=0.0$ and $J=0.0$.
}
\label{figL}
\end{figure}

\begin{figure}
\caption{
%(Color online)
The time dependence of $q$-moments of (a) $\mu_q(t)$ and (b) $\sigma_q(t)^2$,
and those of normal moments of (c) $\mu(t)$ and (d) $\sigma(t)^2$
with $\alpha=0.2$ (chain curves), $\alpha=0.5$ (dashed curves)
and $\alpha=0.8$ (solid curves)
for an input given by
$I(t) = \Delta I \:\Theta(t-2)\Theta(6-t)$ with
$\Delta I=1.0$, $\lambda=1.0$, $\beta=0.5$ and $J=0.0$.
The vertical scale of (b) is different from those of (a), (c) and (d).
}
\label{figP}
\end{figure}

\begin{figure}
\caption{
The $\tau_m$ dependence of 
(a) $S_q$ and (b) $g_q$ with $\tau_m=0.0$ and 
the $\tau_m$ dependence of 
(c) $S_q$ and (d) $g_q$ with $\tau_a=0.0$:
$I=0.0$ (chain curves), $I=0.5$ (dashed curves)
and $I=1.0$ (solid curves) with $\lambda=1.0$, 
$\alpha=0.5$ and $\beta=0.5$.
}
\label{figQ}
\end{figure}

\begin{figure}
\caption{
(Color online)
The $q$ dependence of 
(a) $v_{11}/\sigma_q^2$ ($=1/g_{11} \sigma_q^2$)(chain curve)
and $1/\tilde{g}_{11} \sigma_q^2$ (solid curve),
and (b) $v_{22}/\sigma_q^4$ (chain curve),
$1/\tilde{g}_{22} \sigma_q^4$ (solid curve) and 
$1/g_{22} \sigma_q^4$ (dashed curve):
$g_{ij}$ and $\tilde{g}_{ij}$ are elements of
the generalized and extended Fisher information matrices, 
respectively.
}
\label{figN}
\end{figure}


\newpage
\begin{thebibliography}{99}

\bibitem{Frieden98}B. R. Frieden,
{\it Physics from Fisher information: a unification}
(Cambridge Univ. Press, Cambridge, 1998).

%BG thoery of S(t)
\bibitem{Daems99}D. Daems and G. Nicolis,
Phys. Rev. E {\bf 59}, 4000 (1999).


\bibitem{Bag01}B. C. Bag, S. K. Banik, and D. S. Ray,
Phys. Rev. E {\bf 64}, 026110 (2001).

\bibitem{Bag02}B. C. Bag,
Phys. Rev. E {\bf 66}, 026122 (2002).

\bibitem{Gaonzaler03}D. O. Gonzaler, M. Mayorga. J. Orozco,
and L. R. Salazar,
J. Chem. Phys. {\bf 118}, 6989 (2003).

\bibitem{Ai03}B. Q. Ai, X. J. Wang, G. T. Liu, and L.G. Liu,
Phys. Rev. E {\bf 67}, 022903 (2003).

\bibitem{Goswami05}G. Goswami, B. Mukherjee, and B. C. Bag,
Chem. Phys. {\bf 312}, 47 (2005).

%information geometry
\bibitem{Amari00}S. Amari and H. Nagaoka,
{Methods of Information Geometry}, (AMS and Oxford Univeristy press,
2000).

%Tasallis entropy
\bibitem{Tsallis88}C. Tsallis:
J. Stat. Phys. {\bf 52}, 479 (1988).

\bibitem{Tsallis98}C. Tsallis, R. S. Mendes, 
and A. R. Plastino: Physica A {\bf 261}, 534 (1998). 

\bibitem{Tsallis04}C. Tsallis:
Physica D {\bf 193}, 3 (2004).

\bibitem{Nonext}Lists of many applications of the nonextensive
statistics are available at 
URL: (http://tsallis.cat.cbpf.br/biblio.htm)


%generalized Fisher entropy
\bibitem{Plastino95}A. Palstino and A. R. Plastino,
Physica A {\bf 222}, 347 (1995).
% reaction-diffusion

\bibitem{Tsallis95}C. Tsallis and D. J. Bukman,
Phys. Rev. E {\bf 54}, R2197 (1996).
% reaction-diffusion

\bibitem{Plastino97}A. Palstino, A. R. Plastino, and H. G. Miller,
Physica A {\bf 235}, 577 (1997).
% Tissalis, Fisher

\bibitem{Pennini98}F. Pennini, A. R. Plastino, and A. Plastino,
Physica A {\bf 258}, 446 (1998).
% I = E_q[dln p dln p] Tsallis Fisher

\bibitem{Borland99}L. Borland, F. Pennini, A. R. Plastino,
and A. Plastino,
Eur. Phys. J. B. {\bf 12}, 285 (1999).
% reaction-diffusion

\bibitem{Plastino00}A. R. Plastino, M. Casas, and A. Plastino,
Physica A {\bf 280}, 289 (2000).
% MEM reaction-diffusion

\bibitem{Abe03}S. Abe,
Phys. Rev. E {\bf 68}, 031101 (2003).

\bibitem{Naudts04}J. Naudts, 
J. Ineq. Pure Appl. Math. {\bf 5}, 102 (2004).
%arXiv:math-ph/0402005.
%\bibitem{Naudts05}J. Naudts, 
%Open Sys. \& Information Dyn. {\bf 12}, 13 (2005).

\bibitem{Pennini04}F. Pennini and A. Plastino,
Physica A {\bf 334}, 132 (2004).
% I= E_q[dln P_q d lnP_q]

\bibitem{Portesi06}M. Portesi, A. Plastino, and F. Pennini,
Physica A {\bf 365}, 173 (2006).

\bibitem{Portesi07}M. Portesi, F. Pennini and A. Plastino,
Physica A {\bf 373}, 273 (2007).
% g_{ij}= q E[dln p dln p]

\bibitem{Masi06}M. Masi, arXiv:cond-mat/0611300.


\bibitem{Csiszar72}I. Csisz\'{a}r, 
Periodica Math. Hungar. {\bf 2}, 191 (1972).

\bibitem{Kullback59}S. Kullback,
{\it Information Theory and Statistics}, 
(J. Wiley, New York, 1975).

\bibitem{HHasegawa}Hiroshi Hasegawa, 
Prog. Theor. Phys. Suppl. {\bf 162}, 183 (2006).

\bibitem{Ohara07}A. Ohara, Phys. Lett. A {\bf 370}, 184 (2007).

%review
\bibitem{Lindner04}B. Lindner, J. Garc\'{i}a-Ojalvo, A. Neiman, and
Schimansky-Geil\'{i}er,
Phys. Rep. {\bf 392}, 321 (2004).


\bibitem{Hasegawa07b}H. Hasegawa, Physica A {\bf 374}, 585 (2007).

%\bibitem{AMM}H. Hasegawa, Phys. Rev. E {\bf 67}, 041903 (2003).

\bibitem{Hasegawa03}H. Hasegawa, Phys. Rev. E {\bf 67}, 041903 (2003).

\bibitem{Hasegawa07a}H. Hasegawa,
J. Phys. Soc. Jpn. {\bf 75}, 033001 (2007).

\bibitem{Hasegawa08}H. Hasegawa, 
Physica A (in press) [E-print: arXiv.0708.2563].

\bibitem{Sakaguchi01}H. Sakaguchi:
J. Phys. Soc. Jpn. {\bf 70}, 3247 (2001).

\bibitem{Anten02}C. Anteneodo and C. Tsallis:
J. Math. Phys. {\bf 44}, 5194 (2003).

%\bibitem{Note1}
%Alternatively, when adopt $\theta=I$ in Eq. (\ref{eq:H7}), we get
%$g_q^{(1)} = 
%q\left( E\left[\left( \partial Y(x)/\partial I\right)^2 \right]
%-E\left[ \left( \partial Y(x)/\partial I \right) \right]^2 \right)$,
%where $E[ \cdot]$ expresses the average over $p(x)$ [Eq. (\ref{eq:G25})].
%In our model calculations reported in Secs. 2.5 and 2.6, 
%we have employed the generalized Fisher entropy $g_q$ given 
%by Eq. (\ref{eq:H11}), because its expression is valid also for the MEM.


%\HH
\bibitem{Hasegawa06}H. Hasegawa, 
Physica A {\bf 365}, 383 (2006).
%\bibitem{Note1}Eliminating $c_q$ from 
%$S_q^{(1)}=(1-c_q)/(q-1)$ and $S_q^{(N)}=(1-c_q^N)/(q-1)$,
%we get $(q-1) S_q^{(N)}=1-[1-(q-1)S_q^{(1)}]^N$, which leads to
%Eq. (\ref{eq:H5}).


\bibitem{Fa03}K. S. Fa,
Chem. Phys. {\bf 287}, 1 (2003).

\bibitem{Abramo}M. Abramowitz and I. A. Stegun,
{\it Handbook of Mathematical Functions},
(Dover, New York, 1972).


\bibitem{Hasegawa07d}H. Hasegawa, 
Phys. Rev E {\bf 75}, 051904 (2007).

\bibitem{Hasegawa07c}H. Hasegawa, 
in {\it Neuronal Network Research Horizons}, 
edited by M. L. Weiss 
(Nova Science Publishers, New York, 2007), pp 61.

\end{thebibliography}
\end{document}